\documentclass[conference]{IEEEtran}

\usepackage[utf8]{inputenc}
\usepackage{hyperref}
\usepackage[svgnames]{xcolor}
\usepackage{soul}   
\usepackage{amsmath}
\usepackage{fontawesome}
\usepackage{lipsum} 

\usepackage{multirow}

\usepackage{titlesec}
\titleformat{\section}{\bfseries\large}{\thesection}{1em}{}
\titleformat{\subsection}{\bfseries\normalsize}{\thesubsection}{1em}{}
\titleformat{\subsubsection}{\bfseries\normalsize}{\thesubsubsection}{1em}{}

\usepackage{caption}
\usepackage{amssymb}

\usepackage{booktabs}
\usepackage[normalem]{ulem}
\useunder{\uline}{\ul}{}

\DeclareMathOperator{\KL}{\textit{KL}}
\DeclareMathOperator{\FFN}{\textit{FFN}}
\DeclareMathOperator{\TML}{\textit{TML}}
\DeclareMathOperator*{\argmax}{arg\,max}
\DeclareMathOperator{\Vol}{\textit{Vol}}
\DeclareMathOperator{\L1}{\textit{L1}}

\usepackage{todonotes}

\usepackage[capitalise]{cleveref}

\title{Deep Generative Model-Based Generation of Synthetic Individual-Specific Brain MRI Segmentations}

\author{
    \centering
    \IEEEauthorblockN{
        Ruijie Wang\textsuperscript{1}\IEEEauthorrefmark{1},
        Luca Rossetto\textsuperscript{1,2},
        Susan Mérillat\textsuperscript{3,4},
        Christina Röcke\textsuperscript{3,4},
        Mike Martin\textsuperscript{3,4},
        Abraham Bernstein\textsuperscript{1}
    }
\IEEEauthorblockA{
    \textsuperscript{1}Department of Informatics, University of Zurich, Zurich, Switzerland
    }
\IEEEauthorblockA{
    \textsuperscript{2}School of Computing, Dublin City University, Dublin, Ireland
    }
\IEEEauthorblockA{
    \textsuperscript{3}UZH Healthy Longevity Center, University of Zurich, Zurich, Switzerland
}
\IEEEauthorblockA{
    \textsuperscript{4}Department of Psychology, University of Zurich, Zurich, Switzerland
    }
\IEEEauthorblockA{
    \{ruijie, bernstein\}@ifi.uzh.ch, luca.rossetto@dcu.ie, \{susan.merillat, christina.roecke\}@uzh.ch, m.martin@psychologie.uzh.ch
    }
}

\begin{document}


\maketitle

\begin{abstract}


To the best of our knowledge, all existing methods that can generate synthetic brain magnetic resonance imaging (MRI) scans for a specific individual require detailed structural or volumetric information about the individual's brain.
However, such brain information is often scarce, expensive, and difficult to obtain.
In this paper, we propose the first approach capable of generating synthetic brain MRI segmentations---specifically, 3D white matter (WM), gray matter (GM), and cerebrospinal fluid (CSF) segmentations---for individuals using their easily obtainable and often readily available demographic, interview, and cognitive test information.
Our approach features a novel deep generative model, CSegSynth, which outperforms existing prominent generative models, including conditional variational autoencoder (C-VAE), conditional generative adversarial network (C-GAN), and conditional latent diffusion model (C-LDM).
We demonstrate the high quality of our synthetic segmentations through extensive evaluations.
Also, in assessing the effectiveness of the individual-specific generation, we achieve superior volume prediction, with mean absolute errors of only 36.44\,mL, 29.20\,mL, and 35.51\,mL between the ground-truth WM, GM, and CSF volumes of test individuals and those volumes predicted based on generated individual-specific segmentations, respectively.

\end{abstract}

\section{Introduction}
\label{sec:introduction}

In recent years, the structures and volumes of white matter (WM), gray matter (GM), and cerebrospinal fluid (CSF) in the human brain have garnered significant attention in neuroscience and cognitive science research~\cite{gencc2018diffusion,jiang2023white,thompson2020tracking}.
These three regions are primarily assessed and quantified through the segmentation of brain magnetic resonance imaging (MRI) scans~\cite{zhang2001segmentation}.
However, due to high costs~\cite{ibrahim2012cost} and privacy concerns~\cite{white2022data} associated with acquiring brain MRI scans, the WM, GM, and CSF segmentations are substantially underrepresented in research datasets, leading to a data scarcity issue that hinders many important studies.
Based on machine learning and generative AI techniques, recent efforts for addressing this issue can be broadly categorized into two main types:
First, unconditional data augmentation methods~\cite{DBLP:conf/miccai/KwonHK19,DBLP:journals/cmpb/LiuDBSYB23,DBLP:journals/titb/SunCXGYB22} generate synthetic MRI scans that resemble real scans from their training data but do not correspond to any specific individuals.
Second, conditional individual-specific methods~\cite{DBLP:journals/tmi/YuZWSFB19,DBLP:journals/mia/FernandezPBGTVC24,DBLP:conf/miccai/PinayaTDCFNOC22} aim to generate MRI scans for real-world individuals using detailed structural or volumetric information of the individuals' brain.
However, such required brain information, including precise regional volumes and even neuroimaging scans in other modalities, is often scarce and difficult to obtain.

\begin{figure*}[t]
    \centering
    \includegraphics[width=0.95\linewidth]{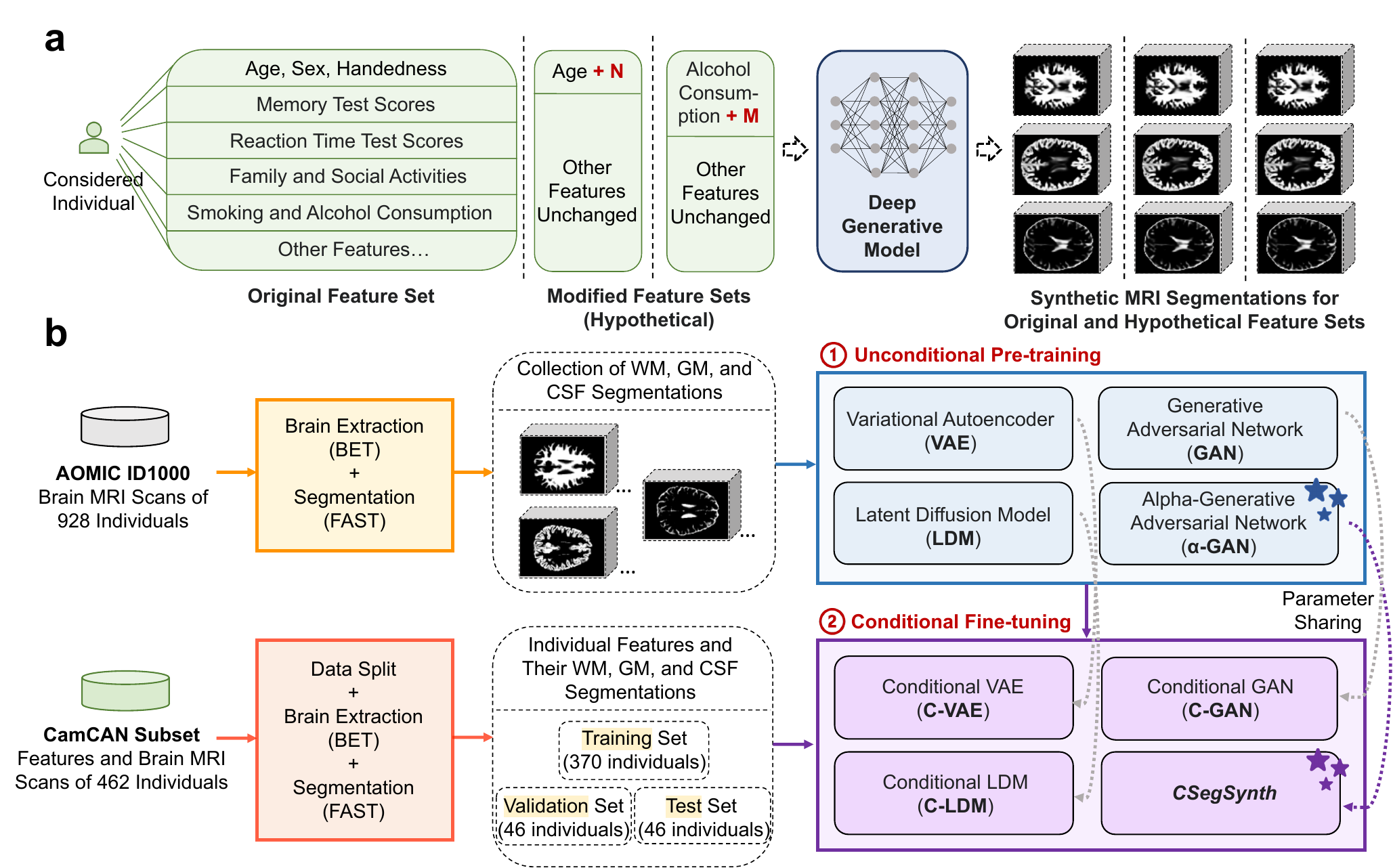}
    \caption{\raggedright \textbf{(a) An overview of how our proposed deep generative model conditionally generates individual-specific MRI segmentations.} Given an individual with easily obtainable features (original or hypothetically modified for studies with control variables), we aim to develop a deep generative model that can generate synthetic 3D MRI segmentations specific to this individual. \textbf{(b) An overview of the proposed approach for training the deep generative model.} The training includes two steps: unconditional pre-training based on AOMIC ID1000~\cite{snoek2021amsterdam} and conditional fine-tuning based on a subset of CamCAN~\cite{shafto2014cambridge}.
    The newly proposed CSegSynth model and its corresponding pre-trained $\alpha$-GAN model are highlighed with stars.} 
    \label{fig:introduction}
\end{figure*}

In this paper, we propose the first approach capable of conditionally generating synthetic individual-specific 3D brain MRI segmentations based on individuals' easily obtainable and often readily available features that are potentially relevant to their brain structures.
As depicted in \cref{fig:introduction}~(a), the features include demographic information (e.g., age and sex), home interview data (e.g., family and social activities), and cognitive test results (e.g., memory test and reaction time test scores).
We aim to develop a deep generative model that can take these features as input and correspondingly generate synthetic 3D WM, GM, and CSF segmentations for the considered individual.
Moreover, the model is expected to make reasonable predictions when the input features are modified (e.g., when adjusting age or alcohol consumption data for aging- or alcohol-related studies).
\emph{To the best of our knowledge, no existing research addresses these needs, leaving a significant gap in the research field.}

An overview of our approach is outlined in \cref{fig:introduction}~(b), where we train deep generative models, including the proposed \emph{\textbf{C}}onditional \emph{\textbf{Seg}}mentation \emph{\textbf{Synth}}esis model---\emph{\textbf{CSegSynth}}, through two steps: \emph{unconditional pre-training} and \emph{conditional fine-tuning}.
In \emph{unconditional pre-training}, we utilize a relatively large collection of publicly available MRI scans, i.e., AOMIC ID 1000~\cite{snoek2021amsterdam}, to train four unconditional generative models: a variational
autoencoder (VAE)~\cite{DBLP:journals/corr/KingmaW13}, a generative
adversarial network (GAN)~\cite{DBLP:conf/nips/GulrajaniAADC17}, a latent diffusion model (LDM)~\cite{DBLP:conf/cvpr/RombachBLEO22}, and an alpha-generative
adversarial network ($\alpha$-GAN)~\cite{DBLP:journals/corr/RoscaLWM17}.
VAE, GAN, and LDM represent three prominent paradigms of generative modeling: variational inference-based, adversarial learning-based, and denoising diffusion-based~\cite{DBLP:journals/pami/Bond-TaylorLLW22}, respectively.
$\alpha$-GAN can be considered as a combination of VAE and GAN, exhibiting strong performance in unconditional MRI generation~\cite{DBLP:journals/corr/RoscaLWM17}.
This first step enables these models to learn detailed structures of the human brain.
After training, they can transform randomly sampled noise into synthetic 3D WM, GM, and CSF segmentations that appear realistic but do not correspond to specific individuals.
Then, in the \emph{conditional fine-tuning} step, we adapt the four models into four corresponding conditional variants: conditional VAE (C-VAE)~\cite{DBLP:conf/nips/SohnLY15}, conditional GAN (C-GAN)~\cite{DBLP:journals/corr/MirzaO14}, conditional LDM (C-LDM)~\cite{DBLP:conf/cvpr/RombachBLEO22}, and CSegSynth.
C-VAE, C-GAN, and C-LDM are implemented following their general frameworks.
CSegSynth has a novel architecture with neural network-based conditioning modules and additional loss terms that we propose based on $\alpha$-GAN (elaborated in \cref{sec:methods}).
Before fine-tuning, the four conditional models inherit most pre-trained parameters of their unconditional variants, as indicated by the dotted arrows labeled ``parameter sharing'' in \cref{fig:introduction}~(b).
Furthermore, we utilize a subset of CamCAN~\cite{shafto2014cambridge} and split it into training, validation, and test sets for model training and evaluations (elaborated in \cref{subsec:experimental_setup}).

\begin{figure*}
    \centering
    \includegraphics[clip, width=0.9\linewidth]{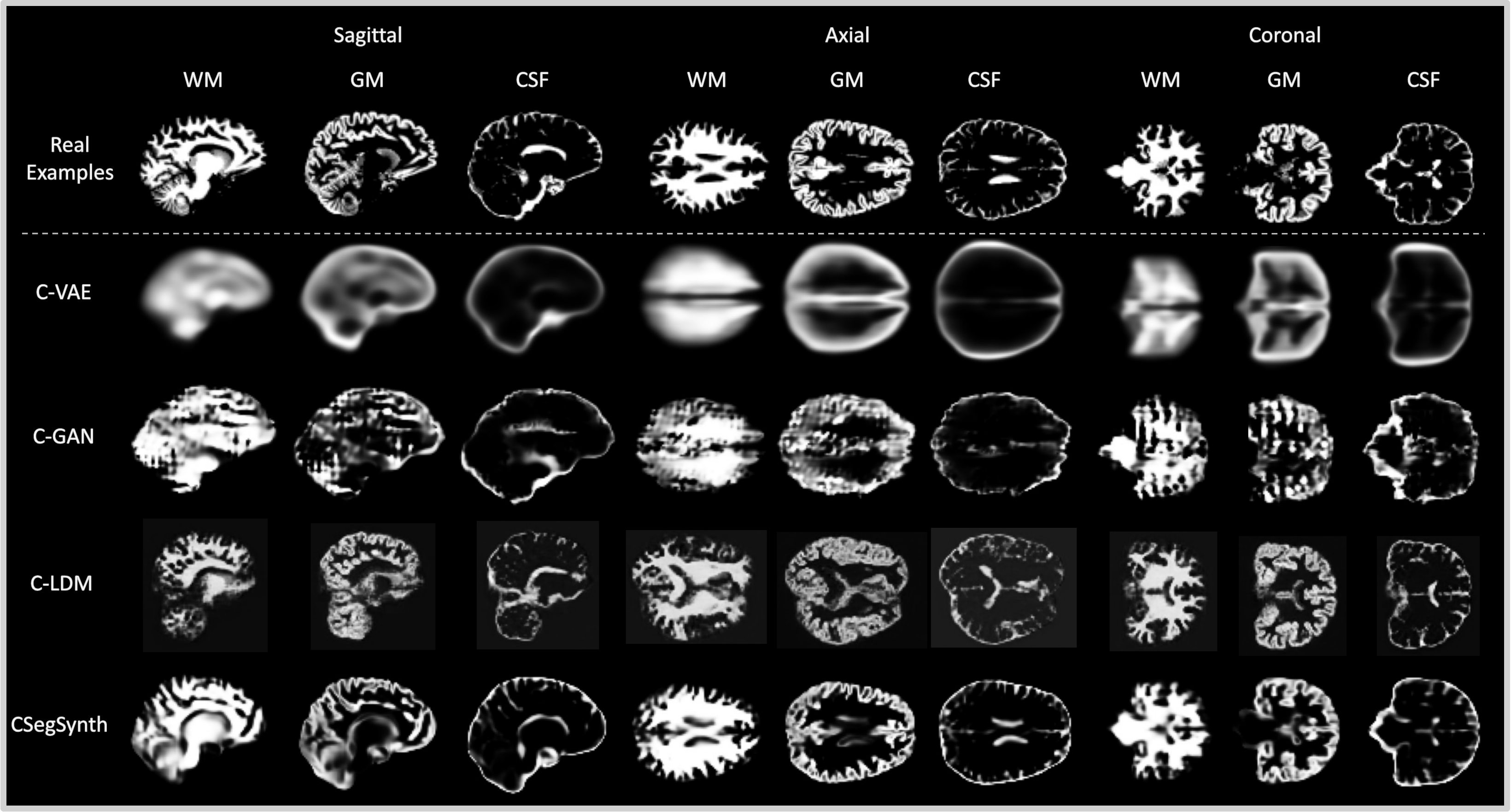}
    \caption{\raggedright \textbf{A comparison of real example segmentations and synthetic segmentations generated by our trained conditional models---C-VAE, C-GAN, C-LDM, and CSegSynth.}
    We present center-cut 2D slices in the sagittal, axial, and coronal views for each 3D segmentation.
    Please note that, due to CamCAN's data restrictions, the real examples are sourced from a different public dataset (AOMIC) and are included solely to illustrate the general image quality of real segmentations.
    }
    \label{fig:mri_quality}
\end{figure*}

Our \emph{main findings} in this paper are as follows:\\
1) The proposed CSegSynth model can generate synthetic 3D brain MRI segmentations with state-of-the-art quality, as demonstrated through qualitative and quantitative evaluations assessing visual appearance, real-versus-synthetic data distribution alignment, image inception distance, and data diversity, in comparison with prominent baseline models, i.e., C-VAE, C-GAN, and C-LDM.
(Reported in \cref{subsec:quality_assessment})\\
2) Based on our proposed training approach in \cref{fig:introduction}~(b), the generation process of C-VAE, C-GAN, and CSegSynth can be effectively conditioned on the input individual features, as demonstrated by evaluations focusing on the accuracy of predicted total WM, GM, and CSF volumes in generated individual-specific segmentations.
In the evaluations, CSegSynth substantially outperforms other deep generative models and several widely-used conventional regression methods that predict these volumes directly based on individual features, including linear regression (LR), polynomial regression (PR), support vector regression (SVR), and feedforward neural network (FFN).
Notably, CSegSynth achieved mean absolute errors of only 36.44\,mL, 29.20\,mL, and 35.51\,mL for total WM, GM, and CSF volumes, respectively.
(Reported in \cref{subsec:conditioning_assessment})\\
3) By computing gradients of generated segmentations with respect to input features, CSegSynth can serve as a convenient testbed for quantifying the impact of input features on human brain structures, offering great potential in neuroscience and cognitive science studies.
Moreover, we probed CSegSynth with a set of hypothetically modified input features with increasing ages in two scenarios: ``ideal'' where other features remain unchanged, and ``regressed'' where other features are adjusted according to age-based regression. CSegSynth generated plausible hypothetical segmentations in both scenarios.
Although we cannot verify the predictions due to the absence of corresponding real brain scans, these observations, along with CSegSynth's performance in all the above evaluations, suggest its promising potential in future studies that focus on human brain structure prediction.
(Discussed in \cref{sec:discussion})

\section{Results}
\label{sec:results}

\subsection{\textbf{Experimental Setup}}
\label{subsec:experimental_setup}

\textbf{Datasets.} As shown in \cref{fig:introduction}~(b), we used AOMIC ID1000~\cite{snoek2021amsterdam} and a subset of CamCAN CC700~\cite{shafto2014cambridge} for the unconditional pre-training and conditional fine-tuning of deep generative models, respectively.
AOMIC ID1000 includes T1-weighted brain MRI scans of 928 healthy young individuals (19 to 26 years old, male/female ratio 0.9), providing the unconditional models with fundamental knowledge of human brain structures.
The CamCAN subset includes T1-weighted brain MRI scans of 462 individuals in a wider age range (18 to 87 years old, male/female ratio 1.0), enabling our conditional models to generalize across different ages.
Also, for the conditional models, we used 198 features from CamCAN for each individual.
Specifically, these features include the CamCAN home interview data (demographic data, lifestyle variables, physical and social activity etc.), physiological measures (cardiovascular measures, height, and weight), and the cognitive data (Cattell~\cite{cattell1971abilities}, Proverbs~\cite{hodges1994neurological}, RT choice~\cite{taylor2017cambridge}, RT simple~\cite{taylor2017cambridge}, Synsem~\cite{rodd2010functional}, and VSTM colour~\cite{zhang2008discrete}).
The detailed variable list and data descriptions are available at \url{https://camcan-archive.mrc-cbu.cam.ac.uk/dataaccess/index.php}.

\textbf{Data Preprocessing.} For the original MRI scans from both AOMIC and CamCAN, we used the Brain Extraction Tool (BET)(\url{https://fsl.fmrib.ox.ac.uk/fsl/fslwiki/BET}) and FMRIB's Automated Segmentation Tool (FAST)(\url{https://fsl.fmrib.ox.ac.uk/fsl/fslwiki/FAST}) for skull stripping and brain segmentation~\cite{zhang2001segmentation}, resulting in ground-truth 3D WM, GM, and CSF segmentations of the size $80\times128\times128$, where each voxel represents the partial volume estimate of WM, GM, or CSF with a cubic size of $8\,mm^3$.
The CamCAN subset was randomly split into a training set (80\%, 370 individuals, male/female ratio 1.1), a validation set (10\%, 46 individuals, male/female ratio 1.3), and a test set (10\%, 46 individuals, male/female ratio 0.8).
Each individual's input features are represented as a single vector that is standardized based on the mean and standard deviation of the feature values in the training set.
The following results are from the test set of CamCAN, which was never disclosed to the trained models.

\begin{table*}[t]
\centering
\caption{\raggedright \textbf{Quantitative evaluation results of the synthetic WM, GM, and CSF segmentations for all test individuals.}
The evaluation metrics include Maximum Mean Discrepancy (MMD), 2D and 3D Fréchet Inception Distance (2D/3D-FID), and the absolute difference in Structural Similarity Index Measure ($|\Delta SSIM|$) between real and synthetic segmentations.
The best results are shown in bold, while the second-best are underlined.
The arrows next to the metrics indicate that lower values correspond to better performance.
}
\label{tab:mri_quality}
\begin{tabular}{@{}cccc|ccc|ccc|ccc@{}}
\toprule
 & \multicolumn{3}{c}{\colorbox{White}{\textbf{MMD}} $\downarrow$} & \multicolumn{3}{c}{\colorbox{White}{\textbf{2D-FID}} $\downarrow$} & \multicolumn{3}{c}{\colorbox{White}{\textbf{3D-FID}} $\downarrow$} & \multicolumn{3}{c}{\colorbox{White}{$\boldsymbol{|\Delta SSIM|}$ $\downarrow$}} \\ \cmidrule{2-13}
\multicolumn{1}{l}{} & \colorbox{White}{\textbf{WM}} & \colorbox{White}{\textbf{GM}} & \colorbox{White}{\textbf{CSF}} & \colorbox{White}{\textbf{WM}} & \colorbox{White}{\textbf{GM}} & \colorbox{White}{\textbf{CSF}} & \colorbox{White}{\textbf{WM}} & \colorbox{White}{\textbf{GM}} & \colorbox{White}{\textbf{CSF}} & \colorbox{White}{\textbf{WM}} & \colorbox{White}{\textbf{GM}} & \colorbox{White}{\textbf{CSF}} \\ \midrule
C-VAE & \colorbox{White}{0.0529} & \colorbox{White}{\underline{0.0410}} & \colorbox{White}{\underline{0.0436}} & \colorbox{White}{0.2940} & \colorbox{White}{0.2544} & \colorbox{White}{0.2789} & \colorbox{White}{0.1439} & \colorbox{White}{0.1203} & \colorbox{White}{0.1305} & \colorbox{White}{0.1188} & \colorbox{White}{0.0938} & \colorbox{White}{0.0836} \\
C-GAN & \colorbox{White}{\underline{0.0339}} & \colorbox{White}{0.0415} & \colorbox{White}{0.0572} & \colorbox{White}{\underline{0.0424}} & \colorbox{White}{{\underline{0.0470}}} & \colorbox{White}{0.0914} & \colorbox{White}{\textbf{0.0064}} & \colorbox{White}{\underline{0.0041}} & \colorbox{White}{\underline{0.0086}} & \colorbox{White}{\underline{0.0162}} & \colorbox{White}{\underline{0.0252}} & \colorbox{White}{0.0697} \\
C-LDM & \colorbox{White}{0.0982} & \colorbox{White}{0.0680} & \colorbox{White}{0.0632} & \colorbox{White}{0.0742} & \colorbox{White}{0.0768} & \colorbox{White}{\textbf{0.0481}} & \colorbox{White}{0.0210} & \colorbox{White}{0.0164} & \colorbox{White}{0.0265} & \colorbox{White}{0.0964} & \colorbox{White}{0.0781} & \colorbox{White}{\underline{0.0443}} \\
CSegSynth & \colorbox{White}{\textbf{0.0273}} & \colorbox{White}{\textbf{0.0240}} & \colorbox{White}{\textbf{0.0203}} & \colorbox{White}{\textbf{0.0361}} & \colorbox{White}{\textbf{0.0438}} & \colorbox{White}{{\ul 0.0695}} & \colorbox{White}{\underline{0.0130}} & \colorbox{White}{\textbf{0.0020}} & \colorbox{White}{\textbf{0.0046}} & \colorbox{White}{\textbf{0.0051}} & \colorbox{White}{\textbf{0.0204}} & \colorbox{White}{\textbf{0.0439}} \\ \bottomrule
\end{tabular}
\end{table*}

\begin{table*}
\centering
\caption{\raggedright \textbf{Accuracy of the WM, GM, and CSF volume predictions for all test individuals.}
The metrics include Mean Absolute Percentage Error (MAPE), Mean Absolute Error (MAE), and Coefficient of Determination (${R^{2}}$).
In addition to the generative models, we also report results for four regression methods: linear regression (LR), polynomial regression (PR), support vector regression (SVR), and feedforward neural network (FFN).
MAE is reported in milliliters (mL) for ease of interpretation.
The best and second-best results for each metric are highlighted in bold and underlined, respectively.
The arrows next to the metrics indicate the preferred direction for each metric.}
\label{tab:vol_statistics}
\begin{tabular}{@{}cccc|ccc|ccc@{}}
\toprule
& \multicolumn{3}{c}{\colorbox{White}{\textbf{WM}}} & \multicolumn{3}{c}{\colorbox{White}{\textbf{GM}}} & \multicolumn{3}{c}{\colorbox{White}{\textbf{CSF}}} \\ \cmidrule(l){2-10} 
& \colorbox{White}{\textbf{MAPE} $\downarrow$} & \colorbox{White}{\textbf{MAE} $\downarrow$}       & \colorbox{White}{$\boldsymbol{R^{2}}$ $\uparrow$} & \colorbox{White}{\textbf{MAPE} $\downarrow$} & \colorbox{White}{\textbf{MAE} $\downarrow$} & \colorbox{White}{$\boldsymbol{R^{2}}$ $\uparrow$} & \colorbox{White}{\textbf{MAPE} $\downarrow$} & \colorbox{White}{\textbf{MAE} $\downarrow$}& \colorbox{White}{$\boldsymbol{R^{2}}$ $\uparrow$}          \\ \midrule
\colorbox{White}{LR}      & \colorbox{White}{43.30\%}  & \colorbox{White}{231.59}    & \colorbox{White}{-20.57}  & \colorbox{White}{40.60\%}   & \colorbox{White}{228.43}    & \colorbox{White}{-22.72}  & \colorbox{White}{58.02\%}   & \colorbox{White}{154.33}    & \colorbox{White}{-16.32}      \\
\colorbox{White}{PR}    & \colorbox{White}{1843.12\%}    & \colorbox{White}{9680.95}   & \colorbox{White}{-0.0007}  & \colorbox{White}{1092.91\%} & \colorbox{White}{5926.84}   & \colorbox{White}{-18623.23} & \colorbox{White}{2251.73\%} & \colorbox{White}{6430.33}   & \colorbox{White}{-28478.56}   \\
\colorbox{White}{SVR}    & \colorbox{White}{11.38\%}   & \colorbox{White}{59.11}     & \colorbox{White}{-0.0082}  & \colorbox{White}{9.98\%}  & \colorbox{White}{55.91}     & \colorbox{White}{-0.0057}  & \colorbox{White}{16.00\%}  & \colorbox{White}{46.23}     & \colorbox{White}{-0.01}     \\
\colorbox{White}{FFN}   & \colorbox{White}{9.21\%}    & \colorbox{White}{50.29}     & \colorbox{White}{0.20}  & \colorbox{White}{8.22\%}     & \colorbox{White}{47.59}     & \colorbox{White}{0.082}  & \colorbox{White}{\underline{12.62\%}}    & \colorbox{White}{38.42}     & \colorbox{White}{0.23}        \\
\colorbox{White}{C-VAE}  & \colorbox{White}{\underline{8.57\%}}   & \colorbox{White}{\underline{46.30}}     & \colorbox{White}{0.13}   & \colorbox{White}{7.33\%}    & \colorbox{White}{42.77}     & \colorbox{White}{0.36}   & \colorbox{White}{13.97\%}    & \colorbox{White}{38.85}     & \colorbox{White}{\underline{0.39}}        \\
\colorbox{White}{C-GAN}   & \colorbox{White}{8.82\%}  & \colorbox{White}{47.49}     & \colorbox{White}{\underline{0.32}}    & \colorbox{White}{\underline{6.10\%}}   & \colorbox{White}{\underline{34.71}}     & \colorbox{White}{\underline{0.63}}   & \colorbox{White}{12.72\%}    & \colorbox{White}{\underline{37.08}}     & \colorbox{White}{\textbf{0.42}}        \\
\colorbox{White}{C-LDM}  & \colorbox{White}{17.29\%}   & \colorbox{White}{97.81}     & \colorbox{White}{-2.17}   & \colorbox{White}{14.85\%}   & \colorbox{White}{84.07}     & \colorbox{White}{-1.56}   & \colorbox{White}{30.80\%}   & \colorbox{White}{92.69}     & \colorbox{White}{-2.27}       \\
\colorbox{White}{CSegSynth}  & \colorbox{White}{\textbf{6.82\%}} & \colorbox{White}{\textbf{36.44}}     & \colorbox{White}{\textbf{0.57}}   & \colorbox{White}{\textbf{5.03\%}}    & \colorbox{White}{\textbf{29.20}}     & \colorbox{White}{\textbf{0.67}}    & \colorbox{White}{\textbf{12.32\%}}   & \colorbox{White}{\textbf{35.51}}     & \colorbox{White}{0.35}        \\ \bottomrule
\end{tabular}
\end{table*}

\subsection{Quality of Synthetic Segmentations}
\label{subsec:quality_assessment}

We report both qualitative and quantitative evaluations of the synthetic segmentations generated by our final conditional models.

\textbf{Qualitative Evaluation.} 
Following existing work~\cite{DBLP:journals/titb/DorjsembePOX24,DBLP:conf/miccai/PinayaTDCFNOC22,DBLP:conf/miccai/KwonHK19}, we randomly select one individual from the CamCAN test set and present synthetic segmentations that our trained conditional models generated for this individual in \cref{fig:mri_quality}.
We present center-cut 2D slices in the sagittal, axial, and coronal views for clarity.
Since CamCAN is not a public dataset, we cannot show the individual’s real segmentations for direct comparison. 
Nevertheless, as our goal in this evaluation is to assess the overall image quality, we include visually similar real example segmentations from the publicly available AOMIC dataset in \cref{fig:mri_quality} for reference.
We summarize our main findings in the following.
Notably, these findings are not specific to the presented individual but generalize across the entire test set.

\begin{itemize}
    \item C-VAE is only able to capture the overall shape of WM, GM, and CSF.
    Its synthetic segmentations appear substantially blurrier than those of the other models.
    This is an inherent limitation of C-VAE and has been also widely observed in other image generation tasks~\cite{DBLP:journals/csur/JabbarLO22}.
    \item The blurriness issue of C-VAE can often be mitigated by replacing the voxel-wise reconstruction loss with the generator-discriminator mini-max loss of C-GAN.
    However, the training of C-GAN is challenging and prone to instability~\cite{DBLP:journals/csur/JabbarLO22}, particularly in our task, where the training data is significantly limited.
    Consequently, in \cref{fig:mri_quality}, we observe substantial unnatural artifacts in the synthetic segmentations of C-GAN.
    \item As a state-of-the-art deep generative model, C-LDM generates segmentations with superior fidelity compared to C-VAE and C-GAN.
    However, we find that C-LDM is prone to distortions and inconsistencies.
    As observed in the axial view of its generated segmentations, the synthetic left and right cerebral hemispheres appear abnormally asymmetric.
    Similar substantial distortions are also evident in the sagittal view.
    \item CSegSynth jointly utilizes the reconstruction loss of C-VAE and the generator-discriminator mini-max loss of C-GAN. 
    It achieves superior fidelity compared to C-VAE and C-GAN.
    Moreover, compared to C-LDM, CSegSynth exhibits stronger robustness against distortions.
    Most importantly, CSegSynth demonstrates the strongest ability among all trained models in capturing detailed brain structures, as evidenced by more realistic cerebral ventricles and cortical gyri in the synthetic segmentations.
\end{itemize}

\begin{table*}
\centering
\caption{\raggedright \textbf{The correlation between ground-truth and predicted WM, GM, and CSF volumes.}
We report Pearson correlation coefficients ($r$), Spearman's rank correlation coefficients ($\rho$), and Kendall's rank correlation coefficients ($\tau$).
The highest coefficients for each region are in bold, whereas the second-highest are underlined.
\sethlcolor{Beige}
The $p$-values of $\rho$ and $\tau$ are reported, according to which the coefficients are highlighted with a \hl{light-green} background for statistical significance ($p \leq 5\times10^{-2}$) and a \sethlcolor{MistyRose}\hl{pink} background for statistical insignificance ($p > 5\times10^{-2}$).
Please note that we report $r$ solely as a descriptive metric of prediction performance without p-values, since the ground-truth CSF volumes and C-LDM outputs do not satisfy the normality assumption according to Shapiro–Wilk tests~\cite{shapiro1965analysis}.
}
\label{tab:correlations}
\begin{tabular}{@{}cccc|ccc|ccc@{}}
\toprule
 &
  \multicolumn{3}{c}{\colorbox{White}{\textbf{WM}}} &
  \multicolumn{3}{c}{\colorbox{White}{\textbf{GM}}} &
  \multicolumn{3}{c}{\colorbox{White}{\textbf{CSF}}} \\ \cmidrule(l){2-10} 
 &
  \colorbox{White}{$\boldsymbol{r}$ $\uparrow$}&
  \colorbox{White}{$\boldsymbol{\rho}$ $\uparrow$, $\boldsymbol{p}$ $\downarrow$}&
  \colorbox{White}{$\boldsymbol{\tau}$ $\uparrow$, $\boldsymbol{p}$ $\downarrow$}&
  \colorbox{White}{$\boldsymbol{r}$ $\uparrow$}&
  \colorbox{White}{$\boldsymbol{\rho}$ $\uparrow$, $\boldsymbol{p}$ $\downarrow$}&
  \colorbox{White}{$\boldsymbol{\tau}$ $\uparrow$, $\boldsymbol{p}$ $\downarrow$}&
  \colorbox{White}{$\boldsymbol{r}$ $\uparrow$}&
  \colorbox{White}{$\boldsymbol{\rho}$ $\uparrow$, $\boldsymbol{p}$ $\downarrow$}&
  \colorbox{White}{$\boldsymbol{\tau}$ $\uparrow$, $\boldsymbol{p}$ $\downarrow$}\\ \midrule
LR &
  \begin{tabular}[c]{@{}c@{}}\colorbox{White}{0.25}\\ \colorbox{White}{$-$}\end{tabular} &
  \begin{tabular}[c]{@{}c@{}}\colorbox{Beige}{0.32}\\ \colorbox{White}{$3\times 10^{-2}$}\end{tabular} &
  \begin{tabular}[c]{@{}c@{}}\colorbox{Beige}{0.21}\\ \colorbox{White}{$4\times 10^{-2}$}\end{tabular} &
  \begin{tabular}[c]{@{}c@{}}\colorbox{White}{0.21}\\ \colorbox{White}{$-$}\end{tabular} &
  \begin{tabular}[c]{@{}c@{}}\colorbox{Beige}{0.31}\\ \colorbox{White}{$3\times 10^{-2}$}\end{tabular} &
  \begin{tabular}[c]{@{}c@{}}\colorbox{Beige}{0.23}\\ \colorbox{White}{$3\times 10^{-2}$}\end{tabular} &
  \begin{tabular}[c]{@{}c@{}}\colorbox{White}{0.17}\\ \colorbox{White}{$-$}\end{tabular} &
  \begin{tabular}[c]{@{}c@{}}\colorbox{MistyRose}{0.28}\\ \colorbox{White}{$6\times 10^{-2}$}\end{tabular} &
  \begin{tabular}[c]{@{}c@{}}\colorbox{Beige}{0.21}\\ \colorbox{White}{$4\times 10^{-2}$}\end{tabular} \\ \cmidrule(l){1-10} 
PR &
  \begin{tabular}[c]{@{}c@{}}\colorbox{White}{0.24}\\ \colorbox{White}{$-$}\end{tabular} &
  \begin{tabular}[c]{@{}c@{}}\colorbox{Beige}{0.31}\\ \colorbox{White}{$3\times 10^{-2}$}\end{tabular} &
  \begin{tabular}[c]{@{}c@{}}\colorbox{MistyRose}{0.18}\\ \colorbox{White}{$7\times 10^{-2}$}\end{tabular} &
  \begin{tabular}[c]{@{}c@{}}\colorbox{White}{0.18}\\ \colorbox{White}{$-$}\end{tabular} &
  \begin{tabular}[c]{@{}c@{}}\colorbox{MistyRose}{0.21}\\ \colorbox{White}{$2\times 10^{-1}$}\end{tabular} &
  \begin{tabular}[c]{@{}c@{}}\colorbox{MistyRose}{0.15}\\ \colorbox{White}{$1\times 10^{-1}$}\end{tabular} &
  \begin{tabular}[c]{@{}c@{}}\colorbox{White}{0.31}\\ \colorbox{White}{$-$}\end{tabular} &
  \begin{tabular}[c]{@{}c@{}}\colorbox{MistyRose}{0.11}\\ \colorbox{White}{$5\times 10^{-1}$}\end{tabular} &
  \begin{tabular}[c]{@{}c@{}}\colorbox{MistyRose}{0.10}\\ \colorbox{White}{$3\times 10^{-1}$}\end{tabular} \\ \cmidrule(l){1-10} 
SVR &
  \begin{tabular}[c]{@{}c@{}}\colorbox{White}{\underline{0.68}}\\ \colorbox{White}{$-$}\end{tabular} &
  \begin{tabular}[c]{@{}c@{}}\colorbox{Beige}{\underline{0.71}}\\ \colorbox{White}{$3\times 10^{-8}$}\end{tabular} &
  \begin{tabular}[c]{@{}c@{}}\colorbox{Beige}{\underline{0.53}}\\ \colorbox{White}{$2\times 10^{-7}$}\end{tabular} &
  \begin{tabular}[c]{@{}c@{}}\colorbox{White}{0.69}\\ \colorbox{White}{$-$}\end{tabular} &
  \begin{tabular}[c]{@{}c@{}}\colorbox{Beige}{0.74}\\ \colorbox{White}{$3\times 10^{-9}$}\end{tabular} &
  \begin{tabular}[c]{@{}c@{}}\colorbox{Beige}{0.54}\\ \colorbox{White}{$1\times 10^{-7}$}\end{tabular} &
  \begin{tabular}[c]{@{}c@{}}\colorbox{White}{0.45}\\ \colorbox{White}{$-$}\end{tabular} &
  \begin{tabular}[c]{@{}c@{}}\colorbox{Beige}{0.57}\\ \colorbox{White}{$4\times 10^{-5}$}\end{tabular} &
  \begin{tabular}[c]{@{}c@{}}\colorbox{Beige}{0.39}\\ \colorbox{White}{$1\times 10^{-4}$}\end{tabular} \\ \cmidrule(l){1-10} 
FFN &
  \begin{tabular}[c]{@{}c@{}}\colorbox{White}{0.67}\\ \colorbox{White}{$-$}\end{tabular} &
  \begin{tabular}[c]{@{}c@{}}\colorbox{Beige}{0.67}\\ \colorbox{White}{$4\times 10^{-7}$}\end{tabular} &
  \begin{tabular}[c]{@{}c@{}}\colorbox{Beige}{0.51}\\ \colorbox{White}{$6\times 10^{-7}$}\end{tabular} &
  \begin{tabular}[c]{@{}c@{}}\colorbox{White}{0.61}\\ \colorbox{White}{$-$}\end{tabular} &
  \begin{tabular}[c]{@{}c@{}}\colorbox{Beige}{0.62}\\ \colorbox{White}{$4\times 10^{-6}$}\end{tabular} &
  \begin{tabular}[c]{@{}c@{}}\colorbox{Beige}{0.47}\\ \colorbox{White}{$5\times 10^{-6}$}\end{tabular} &
  \begin{tabular}[c]{@{}c@{}}\colorbox{White}{0.58}\\ \colorbox{White}{$-$}\end{tabular} &
  \begin{tabular}[c]{@{}c@{}}\colorbox{Beige}{\underline{0.72}}\\ \colorbox{White}{$2\times 10^{-8}$}\end{tabular} &
  \begin{tabular}[c]{@{}c@{}}\colorbox{Beige}{\underline{0.51}}\\ \colorbox{White}{$7\times 10^{-7}$}\end{tabular} \\ \cmidrule(l){1-10} 
C-VAE &
  \begin{tabular}[c]{@{}c@{}}\colorbox{White}{0.52}\\ \colorbox{White}{$-$}\end{tabular} &
  \begin{tabular}[c]{@{}c@{}}\colorbox{Beige}{0.57}\\ \colorbox{White}{$3\times 10^{-5}$}\end{tabular} &
  \begin{tabular}[c]{@{}c@{}}\colorbox{Beige}{0.41}\\ \colorbox{White}{$6\times 10^{-5}$}\end{tabular} &
  \begin{tabular}[c]{@{}c@{}}\colorbox{White}{0.71}\\ \colorbox{White}{$-$}\end{tabular} &
  \begin{tabular}[c]{@{}c@{}}\colorbox{Beige}{0.73}\\ \colorbox{White}{$1\times 10^{-8}$}\end{tabular} &
  \begin{tabular}[c]{@{}c@{}}\colorbox{Beige}{0.51}\\ \colorbox{White}{$6\times 10^{-7}$}\end{tabular} &
  \begin{tabular}[c]{@{}c@{}}\colorbox{White}{0.63}\\ \colorbox{White}{$-$}\end{tabular} &
  \begin{tabular}[c]{@{}c@{}}\colorbox{Beige}{0.60}\\ \colorbox{White}{$1\times 10^{-5}$}\end{tabular} &
  \begin{tabular}[c]{@{}c@{}}\colorbox{Beige}{0.45}\\ \colorbox{White}{$1\times 10^{-5}$}\end{tabular} \\ \cmidrule(l){1-10} 
C-GAN &
  \begin{tabular}[c]{@{}c@{}}\colorbox{White}{0.60}\\ \colorbox{White}{$-$}\end{tabular} &
  \begin{tabular}[c]{@{}c@{}}\colorbox{Beige}{0.62}\\ \colorbox{White}{$4\times 10^{-6}$}\end{tabular} &
  \begin{tabular}[c]{@{}c@{}}\colorbox{Beige}{0.44}\\ \colorbox{White}{$2\times 10^{-5}$}\end{tabular} &
  \begin{tabular}[c]{@{}c@{}}\colorbox{White}{\underline{0.80}}\\ \colorbox{White}{$-$}\end{tabular} &
  \begin{tabular}[c]{@{}c@{}}\colorbox{Beige}{\underline{0.79}}\\ \colorbox{White}{$5\times 10^{-11}$}\end{tabular} &
  \begin{tabular}[c]{@{}c@{}}\colorbox{Beige}{\underline{0.58}}\\ \colorbox{White}{$2\times 10^{-8}$}\end{tabular} &
  \begin{tabular}[c]{@{}c@{}}\colorbox{White}{\underline{0.65}}\\ \colorbox{White}{$-$}\end{tabular} &
  \begin{tabular}[c]{@{}c@{}}\colorbox{Beige}{0.67}\\ \colorbox{White}{$3\times 10^{-7}$}\end{tabular} &
  \begin{tabular}[c]{@{}c@{}}\colorbox{Beige}{0.48}\\ \colorbox{White}{$2\times 10^{-6}$}\end{tabular} \\ \cmidrule(l){1-10} 
C-LDM &
  \begin{tabular}[c]{@{}c@{}}\colorbox{White}{-0.08}\\ \colorbox{White}{$-$}\end{tabular} &
  \begin{tabular}[c]{@{}c@{}}\colorbox{MistyRose}{-0.13}\\ \colorbox{White}{$4\times 10^{-1}$}\end{tabular} &
  \begin{tabular}[c]{@{}c@{}}\colorbox{MistyRose}{-0.10}\\ \colorbox{White}{$3\times 10^{-1}$}\end{tabular} &
  \begin{tabular}[c]{@{}c@{}}\colorbox{White}{-0.13}\\ \colorbox{White}{$-$}\end{tabular} &
  \begin{tabular}[c]{@{}c@{}}\colorbox{MistyRose}{-0.08}\\ \colorbox{White}{$6\times 10^{-1}$}\end{tabular} &
  \begin{tabular}[c]{@{}c@{}}\colorbox{MistyRose}{-0.05}\\ \colorbox{White}{$6\times 10^{-1}$}\end{tabular} &
  \begin{tabular}[c]{@{}c@{}}\colorbox{White}{-0.04}\\ \colorbox{White}{$-$}\end{tabular} &
  \begin{tabular}[c]{@{}c@{}}\colorbox{MistyRose}{-0.13}\\ \colorbox{White}{$4\times 10^{-1}$}\end{tabular} &
  \begin{tabular}[c]{@{}c@{}}\colorbox{MistyRose}{-0.09}\\ \colorbox{White}{$4\times 10^{-1}$}\end{tabular} \\ \cmidrule(l){1-10} 
CSegSynth &
  \begin{tabular}[c]{@{}c@{}}\colorbox{White}{\textbf{0.80}}\\ \colorbox{White}{$-$}\end{tabular} &
  \begin{tabular}[c]{@{}c@{}}\colorbox{Beige}{\textbf{0.80}}\\ \colorbox{White}{$2\times 10^{-11}$}\end{tabular} &
  \begin{tabular}[c]{@{}c@{}}\colorbox{Beige}{\textbf{0.61}}\\ \colorbox{White}{$2\times 10^{-9}$}\end{tabular} &
  \begin{tabular}[c]{@{}c@{}}\colorbox{White}{\textbf{0.82}}\\ \colorbox{White}{$-$}\end{tabular} &
  \begin{tabular}[c]{@{}c@{}}\colorbox{Beige}{\textbf{0.82}}\\ \colorbox{White}{$3\times 10^{-12}$}\end{tabular} &
  \begin{tabular}[c]{@{}c@{}}\colorbox{Beige}{\textbf{0.64}}\\ \colorbox{White}{$4\times 10^{-10}$}\end{tabular} &
  \begin{tabular}[c]{@{}c@{}}\colorbox{White}{\textbf{0.70}}\\ \colorbox{White}{$-$}\end{tabular} &
  \begin{tabular}[c]{@{}c@{}}\colorbox{Beige}{\textbf{0.76}}\\ \colorbox{White}{$1\times 10^{-9}$}\end{tabular} &
  \begin{tabular}[c]{@{}c@{}}\colorbox{Beige}{\textbf{0.56}}\\ \colorbox{White}{$4\times 10^{-8}$}\end{tabular} \\ \bottomrule
\end{tabular}
\end{table*}

\textbf{Quantitative Evaluation.}
We further quantitatively evaluate the quality of the synthetic segmentations generated for all test individuals considering the Maximum Mean Discrepancy (MMD)~\cite{gretton2012kernel}, 2D and 3D Fréchet Inception Distance (2D/3D-FID)~\cite{heusel2017gans}, and the absolute difference in Structural Similarity Index Measure ($|\Delta SSIM|$)~\cite{wang2004image} between real and synthetic segmentations.
The results, presented in \cref{tab:mri_quality}, lead to the following observations:
\begin{itemize}
    \item MMD is a statistical metric that quantifies the discrepancy between two data distributions. 
    We use the MONAI library~\cite{DBLP:journals/corr/abs-2211-02701} to compute the MMD between the real and synthetic segmentations of all test individuals.
    CSegSynth achieves the lowest MMD across all three regions, indicating that its synthetic distribution aligns most closely with the real-world distribution.
    \item FID measures the distance between real and synthetic segmentations in the feature space of a pre-trained image inception model.
    We adopt the pre-trained RadImageNet ResNet50 (2D)~\cite{mei2022radimagenet} and MedicalNet ResNet50 (3D)~\cite{chen2019med3d} for computing 2D-FID and 3D-FID, respectively.
    2D-FID evaluates the 3D segmentations slice by slice.
    We report the average results of the sagittal, axial, and coronal views.
    3D-FID directly evaluates the 3D segmentations as a whole.
    As reported in \cref{tab:mri_quality}, CSegSynth achieves the lowest 2D-FID for WM and GM and the second lowest for CSF.
    Furthermore, CSegSynth achieves the lowest 3D-FID for GM and CSF and the second-lowest for WM.
    These results demonstrate that, compared to other baselines, CSegSynth's synthetic segmentations are generally the most similar to the real ones from the perspective of pre-trained inception models.
    Nevertheless, please note that the two inception models are not particularly pre-trained with brain MRI segmentations.
    Therefore, their computed FID metrics may diverge from our visual assessment.
    For example, the WM segmentations of CSegSynth are visibly superior to those of C-GAN, which is contrary to the 3D-FID results.
    \item SSIM is a perceptual metric commonly used to assess the similarity between a pair of images. 
    Following recent works~\cite{DBLP:journals/titb/DorjsembePOX24,DBLP:conf/miccai/PinayaTDCFNOC22}, we use it to quantify the respective diversity of real and synthetic brain segmentations.
    Specifically, we randomly sample 500 pairs of real segmentations and compute the mean SSIM between them as a measure of the real-world data diversity, i.e., $SSIM_{real}$.
    Then, we apply the same procedure to a model's synthetic segmentations to measure the diversity of this model's output, i.e., $SSIM_{model}$.
    The absolute difference $|\Delta SSIM| = |SSIM_{real} - SSIM_{model}|$ quantifies the gap between the synthetic data's diversity and the real-world diversity.
    CSegSynth achieves the lowest $|\Delta SSIM|$ by a clear margin, indicating its superior performance in generating segmentations with realistic diversity.
    
\end{itemize}

\begin{table*}[]
\caption{\raggedright \textbf{The top-5 most salient features for WM, GM, and CSF volumes, as identified by the trained CSegSynth model.}
The features are sorted according to the absolute value of their computed gradient.
}
\label{tab:saliency_scores}
\centering
\begin{tabular}{@{}cc|cc|cc@{}}
\toprule
\multicolumn{2}{c|}{\colorbox{White}{WM Volume}} & \multicolumn{2}{c|}{\colorbox{White}{GM Volume}}     & \multicolumn{2}{c}{\colorbox{White}{CSF Volume}}                   \\
\colorbox{White}{Feature}              & \multicolumn{1}{c|}{\colorbox{White}{Gradient}} & \colorbox{White}{Feature}                  & \multicolumn{1}{c|}{\colorbox{White}{Gradient}} & \colorbox{White}{Feature}                                & \colorbox{White}{Gradient} \\ \midrule
\colorbox{White}{Sex (Male:0, Female:1)} & \colorbox{White}{-1134.64} & \colorbox{White}{Sex (Male:0, Female:1)} & \colorbox{White}{-978.44} & \colorbox{White}{Sex (Male:0, Female:1)} & \colorbox{White}{-233.00} \\
\colorbox{White}{Height}                & \colorbox{White}{867.03} & \colorbox{White}{Height}                   & \colorbox{White}{777.48}  & \colorbox{White}{Weight}                                 & \colorbox{White}{171.29}  \\
\colorbox{White}{Weight}                & \colorbox{White}{492.05} & \colorbox{White}{Age}                      & \colorbox{White}{-481.22} & \colorbox{White}{Age}                                    & \colorbox{White}{155.33}  \\
\colorbox{White}{Alcohol Consumption}   & \colorbox{White}{387.00} & \colorbox{White}{Employment (No:0, Yes:1)} & \colorbox{White}{371.31}  & \colorbox{White}{Retired (No:0, Yes:1)}                  & \colorbox{White}{138.58}  \\
\colorbox{White}{Weekly Working Hours}  & \colorbox{White}{329.46} & \colorbox{White}{Weight}                   & \colorbox{White}{345.15}  & \colorbox{White}{VSTM colour Precision (set size 1)} &\colorbox{White}{-136.88} \\ \bottomrule
\end{tabular}
\end{table*}

\subsection{Effectiveness of the Individual-Specific Generation}
\label{subsec:conditioning_assessment}

In this section, we evaluate the accuracy of predicted total WM, GM, and CSF volumes in the synthetic segmentations generated for test individuals, serving as a global indicator of the effectiveness of our individual-specific generation.
Specifically, recall that each voxel in the MRI segmentations represents the partial volume estimate of WM, GM, or CSF with a cubic volume of $8\,mm^3$.
Hence, the ground-truth and predicted volumes for each test individual can be computed by multiplying $8\,mm^3$ with the sum of voxel values in the individual's ground-truth and synthetic segmentations, respectively.
Additionally, we implemented four widely used regression methods as baselines: linear regression (LR), polynomial regression (PR), support vector regression (SVR), and feedforward neural network (FFN).
Similar to the generative models, these regression methods take the same feature vectors as input for each test individual.
However, instead of generating segmentations, they directly predict total WM, GM, and CSF volumes, forming a conventional prediction task.

In \cref{tab:vol_statistics}, we present the Mean Absolute Percentage Error (MAPE), Mean Absolute Error (MAE), and Coefficient of Determination ($R^{2}$) between the ground-truth and predicted volumes.
Furthermore, in \cref{tab:correlations}, we report the Pearson correlation coefficient ($r$), Spearman's rank correlation coefficient ($\rho$), and Kendall's rank correlation coefficient ($\tau$) between the ground-truth and predicted volumes.
From the two tables, we can observe that CSegSynth achieves the best performance across all three regions in terms of MAPE, MAE, $r$, $\rho$, and $\tau$.
It also attains the highest $R^{2}$ for WM and GM.
Additionally, the $p$-values of CSegSynth with respect to $\rho$ and $\tau$ are at least one order of magnitude lower than those of other methods, highlighting the significance of the correlation between the ground truth and its predictions.
These results demonstrate that CSegSynth's generation process is effectively conditioned on the input individual features and reveal a novel application that, to the best of our knowledge, has not been explored in related works: predicting individuals' WM, GM, and CSF volumes by conditionally generating their MRI segmentations.

Notably, C-VAE and C-GAN are also highly competitive compared to conventional regression methods, especially regarding MAPE, MAE, and $R^{2}$.
For example, in predicting GM volumes, C-VAE and C-GAN achieve MAPE values of 7.33\% and 6.10\%, respectively, outperforming all regression methods.
This demonstrates the overall effectiveness of the proposed training approach with unconditional pre-training and conditional fine-tuning.
It is worth mentioning that, even with our training approach, C-LDM is still generally ineffective, as evidenced by relatively high errors in \cref{tab:vol_statistics} and insignificant correlations in \cref{tab:correlations}.
This is due to the difficulty of incorporating volume prediction into its training loss, as the model does not perform complete denoising process and, therefore, does not generate specific segmentations during training.
This is an inherent limitation of C-LDM rather than a shortcoming of our training approach.

\section{Discussion}
\label{sec:discussion}

\begin{figure*}
    \centering
    \includegraphics[width=1.\linewidth]{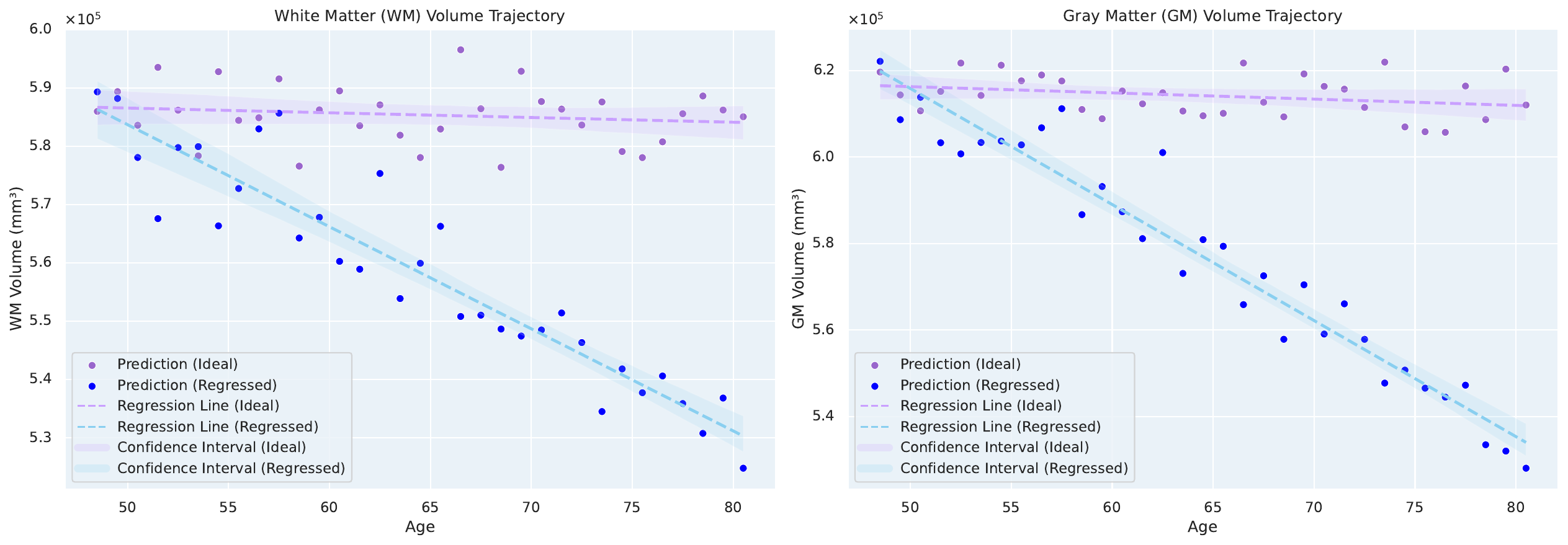}
    \caption{\raggedright \textbf{Predicted trajectories of future WM and GM volumes for a test individual based on future segmentations generated by CSegSynth.}
    We consider two scenarios: ``ideal,'' where all features remain unchanged except for age, and ``regressed,'' where the features are adjusted based on age-related regression.
    For each trajectory, we also plot a linear regression line with a 95\% confidence interval.
    } 
    \label{fig:test_reg_traj}
\end{figure*}

In this section, we discuss two novel applications of CSegSynth and the limitations of this paper with potential directions for future work.

Application-I Quantifying the Impact of Input Features.
Recent neuroscience and cognitive science studies~\cite{heyn2025differential,fletcher2013loss,burgmans2010multiple} have highlighted the significance of quantifying relationships between individual features (e.g., age and cognitive test performance) and human brain structures, including their regional volumes (e.g., WM and GM volumes).
However, this task remains highly challenging due to the hidden, non-linear, and multifaceted nature of these relationships.
As evidenced by accurate volume predictions in \cref{subsec:conditioning_assessment}, CSegSynth effectively models the complex relationships between the adopted individual features and brain regional volumes, offering a promising application based on its differentiable architecture:
When assessing how a particular feature affects the WM, GM, or CSF volume, we can compute the gradient of the volume---derived from synthetic segmentations generated by CSegSynth---with respect to this feature based on the backward pass of the backpropagation algorithm~\cite{rumelhart1986learning}.
The gradient value reflects the extent to which the regional volume would change if this feature varied, thereby quantifying the feature's impact.
Based on the CamCAN test set, the top-5 most salient features for WM, GM, and CSF volumes are presented in \cref{tab:saliency_scores}.
Positive gradients indicate a positive influence, whereas negative gradients indicate a negative influence.
Sex, height, and weight emerge as the top three indicators of WM volume, reflecting the influence of body size: males or individuals with greater height and weight generally exhibit larger WM volumes.
Notably, for GM and CSF volumes, age exerts a relatively stronger influence: its negative gradient for GM and positive gradient for CSF suggest that individuals typically experience decreasing GM and increasing CSF with advancing age.
Working status (e.g., weekly working hours, employment status, and retirement) and alcohol consumption also stand out as salient features---potentially due to their respective correlation with age and sex.
Additionally, the precision (set size 1) metric in the VSTM colour (Visual Short-Term Memory) test~\cite{zhang2008discrete} shows a negative association with CSF volume---higher precision scores are associated with lower expected CSF volumes.
Please note that verifying and rigorously interpreting the computed gradients would require additional data and is beyond the scope of this paper.
Nonetheless, such insights and this procedure could be valuable for brain-related studies, facilitating hypothesis generation and the identification of key features that merit further investigation.
While we do not intend to claim definitive findings, we hope to highlight this promising and novel application, which, to the best of our knowledge, has not been explored in existing synthetic MRI generation works.
Furthermore, the above procedure can be conducted in two more fine-grained cases: 1) computing gradients for individuals within specific subgroups (e.g., subgroups of different ages and sexes) to analyze how the input features differentially influence different individual groups, and 2) computing gradients after extracting certain substructures in the generated segmentations (e.g., cerebral ventricles and cortical gyri) to analyze how the input features influence certain brain substructures.

Application-II Hypothetical Segmentation Generation.
This application of CSegSynth enables the prediction of hypothetical brain segmentations and the creation of a brain digital twin for individuals under controlled developmental conditions~\cite{wang2024virtual,lavanga2023virtual}, such as simulating aging or modeling the effects of working status and daily activities.
In the following, we present a use case where we predict future WM and GM volumes for a middle-aged male test individual from CamCAN.
The exact age cannot be reported due to data privacy concerns.
Specifically, we generate this individual's future feature vectors by increasing the age value under two scenarios: ``ideal'' and ``regressed.''
In the ``ideal'' scenario, all the other features remain unchanged, i.e., assuming this individual is able to maintain all the other conditions despite aging.
In the ``regressed'' scenario, for each of the other features, we train a 3-degree polynomial regression model based on other male individuals in the CamCAN dataset to predict the regressed future value.
We generate future brain segmentations for this individual based on these hypothetical feature vectors.
Then, we compute future WM and GM volumes based on the generated segmentations.
Trajectories of the future WM and GM volumes are plotted in \cref{fig:test_reg_traj}.
We can observe that, if the current conditions (i.e., all other features) can be maintained despite aging, WM and GM volumes are predicted to remain relatively stable over time.
In contrast, if the individual’s other features regress with aging, WM and GM volumes are expected to decline substantially and would align closely with the population trend of the CamCAN dataset, which we computed via linear regression but cannot show here directly due to data restrictions.
Finally, we emphasize that these are only hypothetical observations based on synthetic segmentations. 
Due to the absence of ground-truth future data for this individual, we cannot validate these predictions. 
This analysis is intended solely to demonstrate a potential application of CSegSynth, rather than to establish any claims or findings.

This work has several limitations that require future efforts to address:
First, although CSegSynth has achieved state-of-the-art performance in the generation of high-quality individual-specific segmentations, a noticeable gap remains between the real and generated segmentations, as shown in \cref{fig:mri_quality}.
This limitation is primarily due to the relatively small size of the training dataset.
We will explore how to further improve the quality of generated segmentations by incorporating additional training data.
Second, in \cref{subsec:conditioning_assessment}, the effectiveness of the individual-specific generation is evaluated in terms of overall WM, GM, and CSF volumes.
It could be beneficial to conduct a more fine-grained assessment involving specific brain structures. 
However, as this lies beyond the scope of this paper (i.e., the development of the generative model rather than the downstream analysis), we consider this as an important direction for future work.
Third, in the above two applications, there are extensive observations that align with domain knowledge and intuition.
However, due to the lack of ground-truth data, these observations cannot be verified.
We are actively seeking such datasets with the goal of conducting more rigorous analyses.

\section{Methods}
\label{sec:methods}

In this section, we introduce the details of our approach, including the adopted deep generative models and some implementation details.

\begin{figure*}
    \centering
    \includegraphics[width=0.85\linewidth]{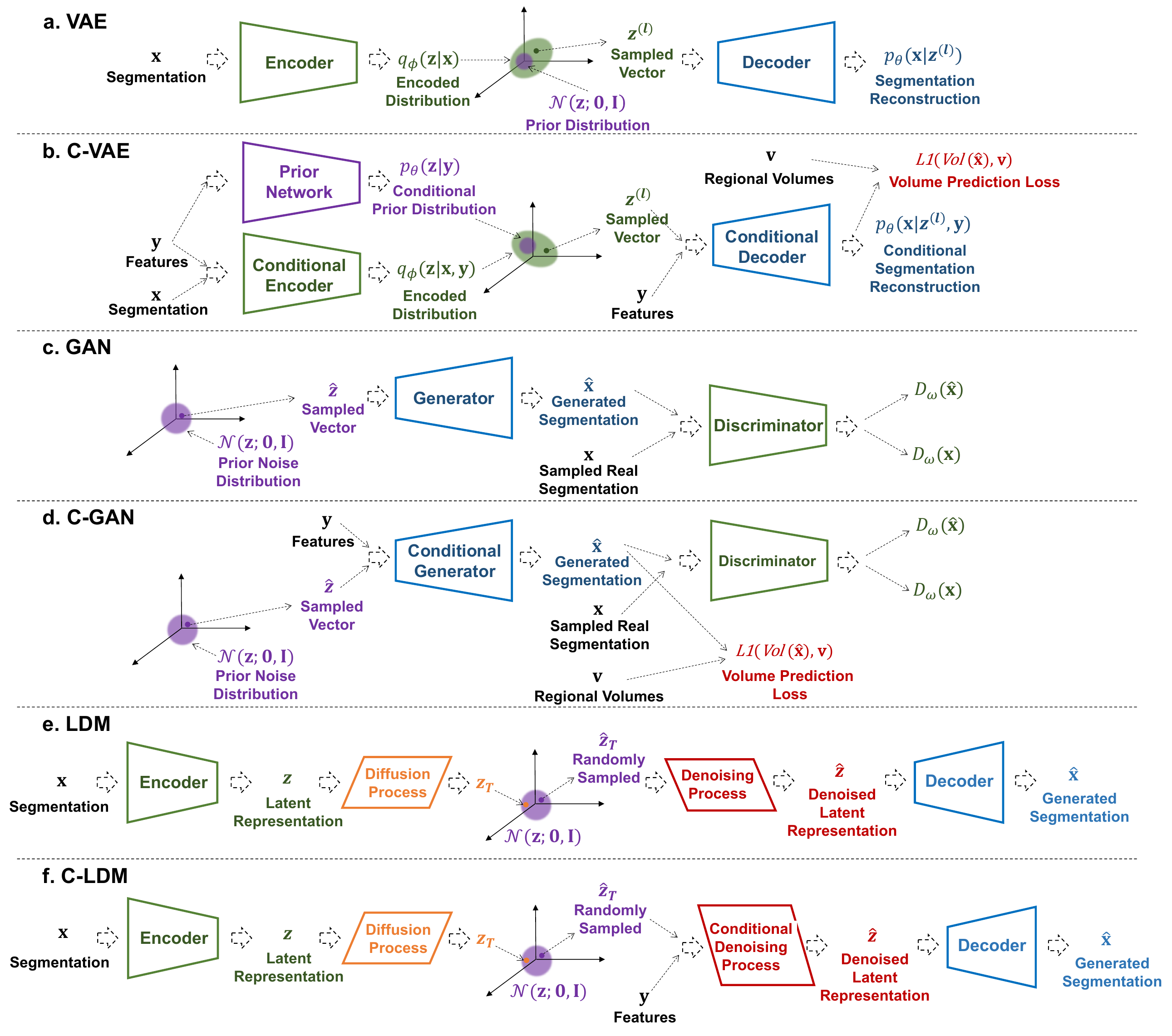}
    \caption{\textbf{An overview of the model structures of VAE, C-VAE, GAN, C-GAN, LDM, and C-LDM.}
    The triplet margin loss of C-GAN is not shown for clarity.
    Please refer to \cref{subsubsec:cgan} for details.
    } 
    \label{fig:structures}
\end{figure*}

\subsection{Notation Definition}

We first define the notations that are necessary for the discussion.
After the brain extraction and segmentation introduced in \cref{subsec:experimental_setup}, the AOMIC dataset~\cite{snoek2021amsterdam} for model pre-training is denoted as $\{(\mathbf{x}^{i}_{wm}, \mathbf{x}^{i}_{gm}, \mathbf{x}^{i}_{csf})\}$, where $\mathbf{x}^{i}_{wm}$, $\mathbf{x}^{i}_{gm}$, and $\mathbf{x}^{i}_{csf}$ are the WM, GM, and CSF segmentations of the $i$-th individual in the dataset.
Each segmentation is a 3-dimensional tensor of the size 80$\times$128$\times$128, i.e., $\mathbf{x}^{i}_{wm}, \mathbf{x}^{i}_{gm}, \mathbf{x}^{i}_{csf}\in [0, 1]^{80\times128\times128}$, where each value in the tensor is between 0 and 1.
For the convenience of computing, regarding each individual, we stack the three segmentations together with the background to obtain a 4-dimensional tensor:
\begin{equation}
    \mathbf{x}^{i} =  \left[\mathbf{x}^{i}_{wm}, \mathbf{x}^{i}_{gm}, \mathbf{x}^{i}_{csf}, (1 - \mathbf{x}^{i}_{wm} - \mathbf{x}^{i}_{gm} - \mathbf{x}^{i}_{csf})\right].
\end{equation}
In this way, the pre-training dataset is denoted as $\mathcal{X}_{pre} = \{\mathbf{x}^{i}\}$, where $\mathbf{x}^{i} \in [0, 1]^{4\times80\times128\times128}$.
The fine-tuning dataset is from the CamCAN repository (\url{http://www.mrc-cbu.cam.ac.uk/datasets/camcan/})~\cite{taylor2017cambridge,shafto2014cambridge}.
It is denoted as $\mathcal{X}_{fin} = \{(\mathbf{x}^{i}, \mathbf{v}^{i}, \mathbf{y}^{i})\}$, where $\mathbf{x}^{i}\in [0, 1]^{4\times80\times128\times128}$ denotes the stacked segmentations and the background, $\mathbf{v}^{i}\in \mathbb{R}^{4}$ is a 4-dimensional vector denoting the ground-truth WM, GM, CSF, and background volumes, respectively, and $\mathbf{y}^{i}$ is a feature vector representing the individual's input features.

\subsection{The Pre-Training Process}

The pre-training process aims to utilize $\mathcal{X}_{pre} = \{\mathbf{x}^{i}\}$ to train four generative models: VAE, GAN, LDM, and $\alpha$-GAN.

\subsubsection{Variational Autoencoder (VAE)}

\begin{figure}[t]
    \centering
    \includegraphics[width=0.75\linewidth]{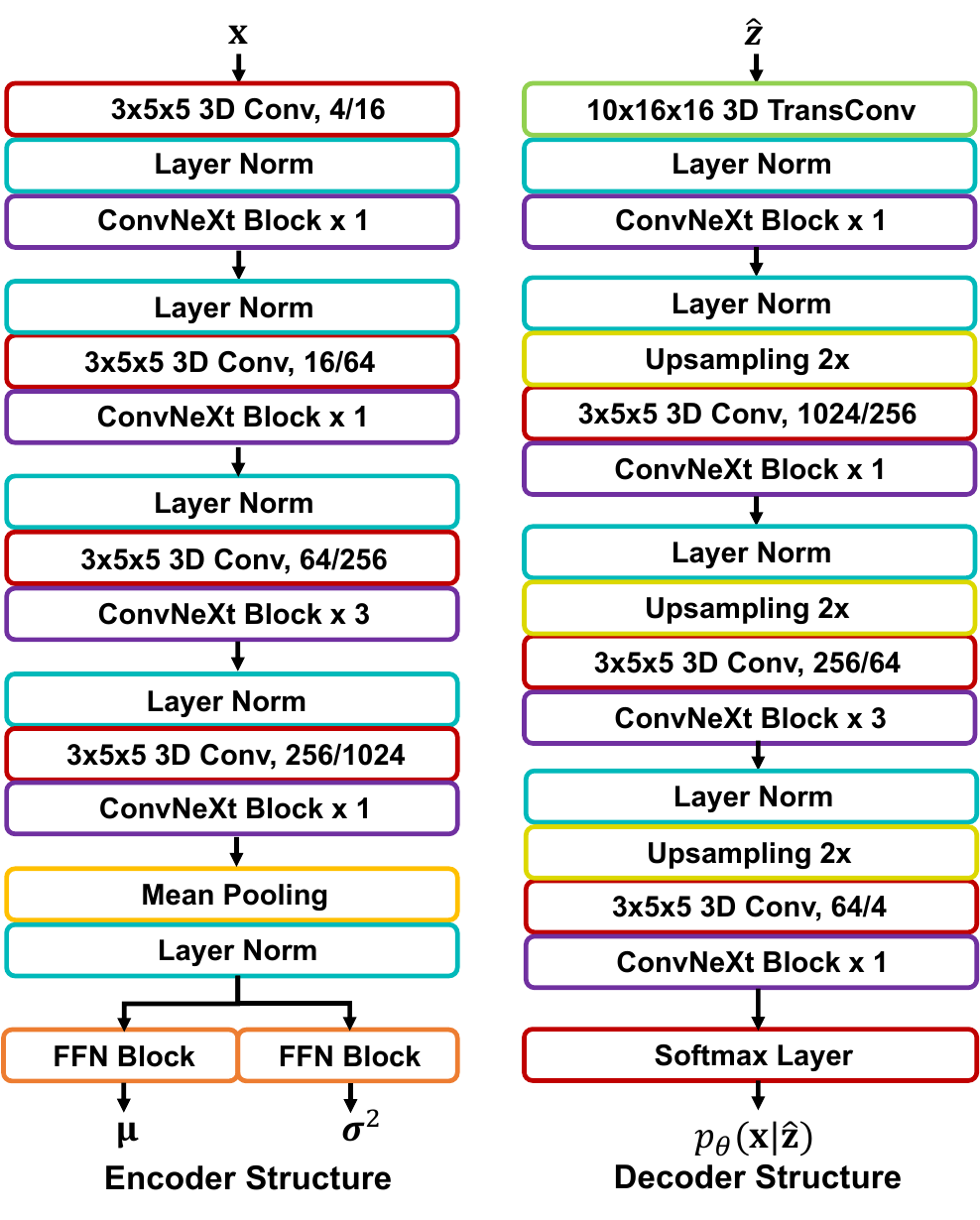}
    \caption{\textbf{An overview of the encoder and decoder structures of the VAE model.} 
    The 3D convolutional (3D Conv) networks are labeled with kernel size (e.g., $3\times5\times5$) and input/output channels (e.g., $4/16$).
    The ConvNeXt~\cite{DBLP:conf/cvpr/0003MWFDX22} blocks are labeled with the repetition times (e.g., $\times 3$).
    The 3D transposed convolutional (3D TransConv) network is labeled with the kernel size $10\times16\times16$.
    The upsampling layers are labeled with the upsampling factor $2\times$.
    Layer Norm and FFN refer to layer normalization~\cite{ba2016layer} and feedforward neural networks, respectively.
    }
    \label{fig:enc_dec}
\end{figure}

An overview of the VAE model~\cite{DBLP:journals/corr/KingmaW13} is depicted in \cref{fig:structures}~(a), where the model consists of an encoder and a decoder.
The encoder encodes each input segmentation $\mathbf{x}\in [0, 1]^{4\times80\times128\times128}$ from $\mathcal{X}_{pre}$ into the distribution $q_{\phi}(\mathbf{z}|\mathbf{x})$, which is assumed to be an isotropic multivariate Gaussian in a 1024-dimensional hidden space, i.e., $q_{\phi}(\mathbf{z}|\mathbf{x}) = \mathcal{N}(\mathbf{z}; \boldsymbol{\mu}, \boldsymbol{\sigma}^{2} \mathbf{I})$, where $\boldsymbol{\mu}, \boldsymbol{\sigma}^{2} \in \mathbb{R}^{1024}$.
Then, a vector $\mathbf{z}^{(l)}$ is sampled from $\mathcal{N}(\mathbf{z}; \boldsymbol{\mu}, \boldsymbol{\sigma}^{2} \mathbf{I})$ and decoded by the decoder to obtain a reconstruction of the input segmentation.
We present an overview of the encoder and decoder structures in \cref{fig:enc_dec}.

The training loss regarding each input segmentation $\mathbf{x}$ is defined as follows:
\begin{equation}
\begin{aligned}
    \mathcal{L}_{vae}&(\mathbf{x};\phi,\theta) = \\& \lambda_1 \KL\left(q_{\phi}\left(\mathbf{z}|\mathbf{x}\right)\|\mathcal{N}\left(\mathbf{z}; \mathbf{0}, \mathbf{I}\right)\right) - \lambda_2 \frac{1}{L} \sum_{l=1}^{L} \log p_{\theta}(\mathbf{x}|\mathbf{z}^{(l)}),
\end{aligned}
\end{equation}
where $\phi$ and $\theta$ denote the parameters of the encoder and decoder, respectively.
$\KL\left(q_{\phi}(\mathbf{z}|\mathbf{x})\|\mathcal{N}(\mathbf{z}; \mathbf{0}, \mathbf{I})\right)$ denotes the Kullback–Leibler (KL) divergence between the encoded distribution and the assumed prior distribution (a standard Gaussian).
$\mathbf{z}^{(l)}$ is the $l$-th sample drawn from $q_{\phi}(\mathbf{z}|\mathbf{x})$ based on the reparameterization trick~\cite{DBLP:journals/corr/KingmaW13}.
$\lambda_{1}$ and $\lambda_{2}$ are hyperparameters for balancing two loss terms.
We employ the AdamW~\cite{DBLP:conf/iclr/LoshchilovH19} optimization algorithm to optimize $\phi$ and $\theta$.
After training, a new segmentation can be generated by decoding a vector that is randomly sampled from the prior distribution.

\subsubsection{Generative Adversarial Network (GAN)}

We adopt the Wasserstein GAN with gradient penalty~\cite{DBLP:conf/nips/GulrajaniAADC17} in our approach.
An overview of the model structure is given in \cref{fig:structures}~(c).
Specifically, the model consists of a generator $G_{\theta}$ and a discriminator $D_{\omega}$ that compete with each other during training: the discriminator aims to discriminate synthetic segmentations generated by the generator from real segmentations, while the generator aims to improve the quality of its synthetic segmentations to fool the discriminator.
The parameters of the generator and the discriminator, i.e., $\theta$ and $\omega$, respectively, are optimized in separate steps with different loss functions.
The loss function for the discriminator is:
\begin{equation}
    \begin{aligned}
     \mathcal{L}_{dis}&(\mathbf{x}, \hat{\mathbf{x}}, \tilde{\mathbf{x}};\omega) = \\& D_{\omega}(\hat{\mathbf{x}}) - D_{\omega}(\mathbf{x}) + \lambda (\|\nabla_{\tilde{\mathbf{x}}}D_{\omega}(\tilde{\mathbf{x}}) \|_{2} - 1)^{2},
\end{aligned}
\end{equation}
where $\mathbf{x}$ is a sampled real training segmentation.
$\hat{\mathbf{x}}$ is a synthetic segmentation generated by the generator based on a noise vector $\hat{\mathbf{z}}\in \mathbb{R}^{1024}$ that is sampled from the prior distribution---a multivariate standard Gaussian $\mathcal{N}(\mathbf{z}; \mathbf{0}, \mathbf{I})$, i.e., $\hat{\mathbf{x}} = G_{\theta}(\hat{\mathbf{z}}), \hat{\mathbf{z}} \sim \mathcal{N}(\mathbf{z}; \mathbf{0}, \mathbf{I})$.
$D_{\omega}(\hat{\mathbf{x}}), D_{\omega}(\mathbf{x})\in \mathbb{R}$ are computed by the discriminator for $\hat{\mathbf{x}}$ and $\mathbf{x}$, respectively.
The term $\lambda (\|\nabla_{\tilde{\mathbf{x}}}D_{\omega}(\tilde{\mathbf{x}}) \|_{2} - 1)^{2}$ is the gradient penalty~\cite{DBLP:conf/nips/GulrajaniAADC17} that aims to enforce the Lipschitz constraint~\cite{DBLP:conf/nips/GulrajaniAADC17} over $D_{\omega}$, where $\tilde{\mathbf{x}} = \epsilon \mathbf{x} + (1 - \epsilon) \hat{\mathbf{x}}$, and $\epsilon$ is sampled from a uniform distribution between 0 and 1, i.e., $\epsilon \sim U\left[0, 1\right]$.
$\lambda$ is a hyperparameter for balancing loss terms.
Then, the loss function for the generator is defined as: 
\begin{equation}
    \mathcal{L}_{gen}(\hat{\mathbf{z}};\theta) = - D_{\omega}(G_{\theta}(\hat{\mathbf{z}})),
\end{equation}
which aims to optimize $\theta$ to improve the discriminator's ``score'' for the generator's synthetic segmentation $G_{\theta}(\hat{\mathbf{z}})$.

In our implementation, the generator and discriminator share the basic structures of the decoder and encoder of the VAE model (cf. \cref{fig:enc_dec}), respectively.
One major difference is that the output layer of the discriminator is only one feedforward neural network (FFN) block that has real-number outputs.
The GAN model is trained over $\mathcal{X}_{pre}$ with the Adam optimization algorithm~\cite{DBLP:journals/corr/KingmaB14}.
After training, we can sample noise vectors from the prior noise distribution and use the generator to generate new synthetic segmentations.

\begin{figure*}
    \centering
    \includegraphics[width=1\linewidth]{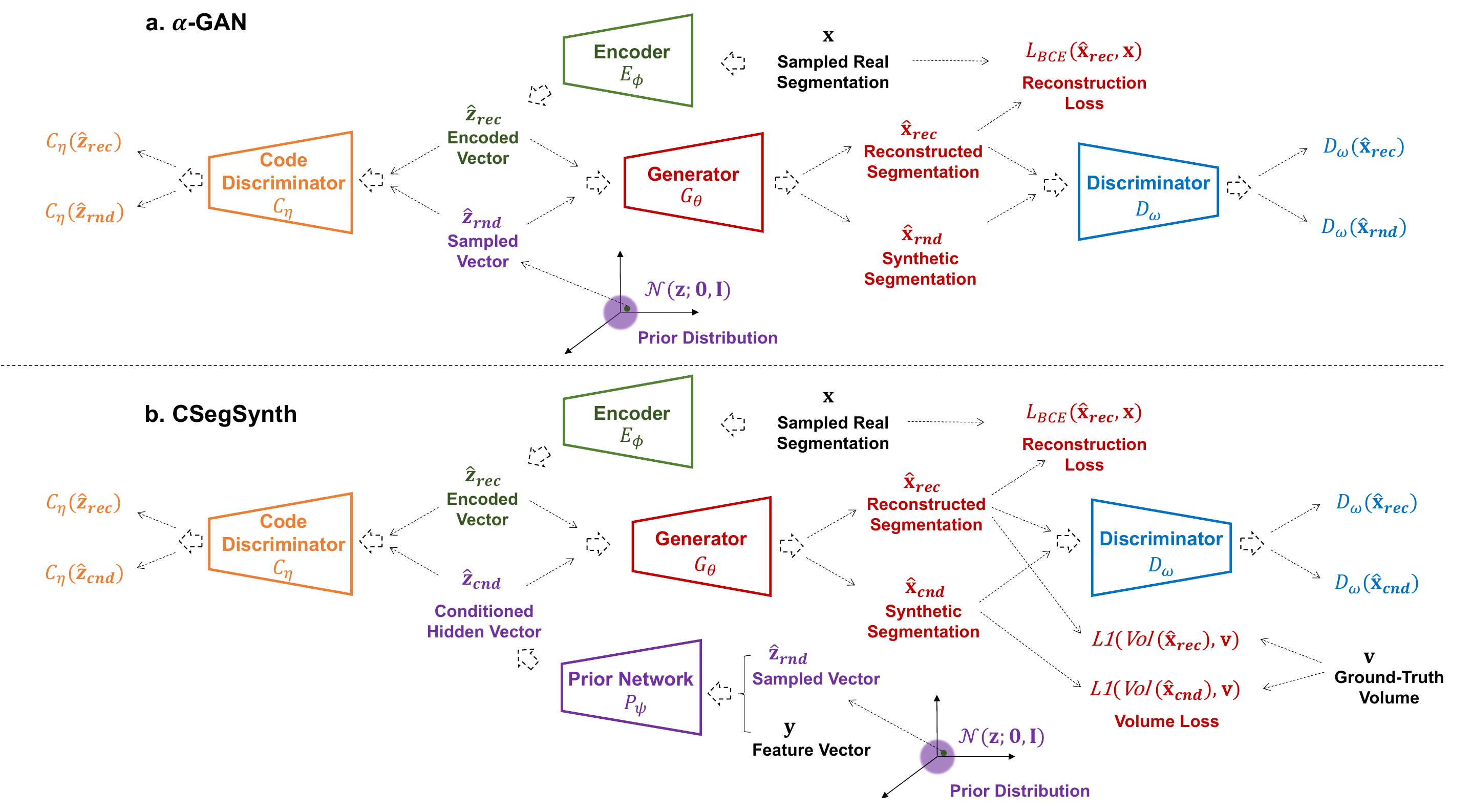}
    \caption{\textbf{An overview of the model structures of $\alpha$-GAN and CSegSynth.}
    The triplet margin loss of CSegSynth is not shown for clarity.
    Please refer to \cref{subsubsec:csegsynth} for details.
    } 
    \label{fig:csegsynth_structures}
\end{figure*}

\subsubsection{Latent Diffusion Model (LDM)}

We adopted the MONAI library~\cite{DBLP:journals/corr/abs-2211-02701} to implement and train an LDM model~\cite{DBLP:conf/cvpr/RombachBLEO22}, which consists of an autoencoder (comprising an encoder and a decoder), a forward diffusion process, and a reverse diffusion (i.e., denoising) process.
An overview of the model is presented in \cref{fig:structures}~(e).
Similar to VAE, the autoencoder represents the original segmentations in a more compact and efficient latent space.
The encoder and decoder are trained with a combination of L1 reconstruction loss, perceptual loss~\cite{zhang2018unreasonable}, adversarial loss~\cite{esser2021taming}, and KL regularization, as proposed by Pinaya et al.~\cite{DBLP:conf/miccai/PinayaTDCFNOC22}.
In the diffusion process, LDM adds noise to the latent representation of the given segmentation $\mathbf{x}$, i.e., $\mathbf{z}$, step by step and eventually converts it into $\mathbf{z_{T}}$, which is close to a sample from the standard Gaussian $\mathcal{N}(\mathbf{z}; \mathbf{0}, \mathbf{I})$.
Here, $T$ denotes the number of steps and is set to 1,000 in our experiment.
In the denoising process, given a latent representation sampled from $\mathcal{N}(\mathbf{z}; \mathbf{0}, \mathbf{I})$, i.e., $\hat{\mathbf{z}}_{T}$, LDM aims to remove the presumed noise in $\hat{\mathbf{z}}_{T}$ step by step and converts it into $\hat{\mathbf{z}}$ that can be decoded into a plausible synthetic segmentation $\hat{\mathbf{x}}$.
The forward diffusion process only requires sampling the noise to add in each step from a standard Gaussian and does not involve training parameters.
The denoising process leverages a UNet module~\cite{ronneberger2015u} to predict the noise to remove at each step.
Its training loss is defined as:
\begin{equation}
\label{equ:ldm}
    \begin{aligned}
        \mathcal{L}_{ldm}(\mathbf{z}, \epsilon, t;\theta) = \|\epsilon - \epsilon_{\theta}(\mathbf{z_{t}}, t)\|_{2},
    \end{aligned}
\end{equation}
where $\mathbf{z}$ is the latent representation of a sampled real training segmentation.
$\epsilon$ is randomly sampled from a standard Gaussian.
$t$ denotes a step number uniformly sampled between 1 and $T$.
$\theta$ denotes the UNet module's parameters.
$\mathbf{z_{t}}$ denotes the result of the diffusion process starting with $\mathbf{z}$ after $t$ steps, which can be obtained based on $\epsilon$ \cite{ho2020denoising}.
$\epsilon_{\theta}(\mathbf{z_{t}}, t)$ denotes the output of the UNet module given $\mathbf{z_{t}}$ and $t$.
The module is trained based on the Adam algorithm~\cite{DBLP:journals/corr/KingmaB14}.
After training, the UNet module can be used to denoise latent representations randomly sampled from $\mathcal{N}(\mathbf{z}; \mathbf{0}, \mathbf{I})$ to generate new segmentations~\cite{ho2020denoising}.

\subsubsection{Alpha-Generative Adversarial Network ($\alpha$-GAN)}

The $\alpha$-GAN model~\cite{DBLP:journals/corr/RoscaLWM17,DBLP:conf/miccai/KwonHK19} consists of an encoder $E_{\phi}$, a code discriminator $C_{\eta}$, a generator $G_{\theta}$, and a discriminator $D_{\omega}$, as depicted in \cref{fig:csegsynth_structures}~(a).
The $\alpha$-GAN model is trained with three loss functions.
The loss function for the encoder and generator is:
\begin{equation}
    \begin{aligned}
        \mathcal{L}_{a1}&(\mathbf{x}, \hat{\mathbf{z}}_{rnd};\phi, \theta) = \\&- D_{\omega}(\hat{\mathbf{x}}_{rnd}) - D_{\omega}(\hat{\mathbf{x}}_{rec}) - \lambda_1 C_{\eta}(\hat{\mathbf{z}}_{rec}) \\&+ \lambda_2 L_{BCE}(\hat{\mathbf{x}}_{rec}, \mathbf{x}),
    \end{aligned}
\end{equation}
where $\mathbf{x}$ is a sampled real segmentation.
$\hat{\mathbf{z}}_{rnd}\sim \mathcal{N}(\mathbf{z};\mathbf{0},\mathbf{I})$ is a random noise sampled from the prior distribution.
$\phi$ and $\theta$ denote the parameters of the encoder and generator, respectively.
$\hat{\mathbf{x}}_{rnd} = G_{\theta}(\hat{\mathbf{z}}_{rnd})$ is a new synthetic segmentation generated by the generator $G_{\theta}$ based on $\hat{\mathbf{z}}_{rnd}$.
$\hat{\mathbf{z}}_{rec} = E_{\phi}(\mathbf{x})$ is the encoded vector of $\mathbf{x}$.
$\hat{\mathbf{x}}_{rec} = G_{\theta}(\hat{\mathbf{z}}_{rec})$ is a reconstruction of $\mathbf{x}$ based on $\hat{\mathbf{z}}_{rec}$.
$C_{\eta}(\hat{\mathbf{z}}_{rec})$ denotes the ``score'' computed by the code discriminator $C_{\eta}$ regarding $\hat{\mathbf{z}}_{rec}$.
$L_{BCE}(\hat{\mathbf{x}}_{rec}, \mathbf{x})$ denotes the voxel-wise binary cross entropy (BCE) loss between the reconstruction $\hat{\mathbf{x}}_{rec}$ and the original segmentation $\mathbf{x}$.

The loss function for training the discriminator $D_{\omega}$ is:
\begin{equation}
    \begin{aligned}
        \mathcal{L}_{a2}&(\mathbf{x}, \hat{\mathbf{z}}_{rnd};\omega) =
        \\& D_{\omega}(\hat{\mathbf{x}}_{rnd}) + D_{\omega}(\hat{\mathbf{x}}_{rec}) 
        - 2 D_{\omega}(\mathbf{x}) 
        \\& + \lambda_3 \left(
        \left(\|\nabla_{\tilde{\mathbf{x}}}D_{\omega}(\tilde{\mathbf{x}}) \|_{2} - 1\right)^{2} +\left(\|\nabla_{\tilde{\mathbf{x}}'}D_{\omega}(\tilde{\mathbf{x}}'\right) \|_{2} - 1)^{2}\right),        
    \end{aligned}
\end{equation}
where $\tilde{\mathbf{x}} = \epsilon \mathbf{x} + (1 - \epsilon) \hat{\mathbf{x}}_{rnd}, \tilde{\mathbf{x}}' = \epsilon' \mathbf{x} + (1 - \epsilon') \hat{\mathbf{x}}_{rec},$ and $\epsilon, \epsilon'\sim U\left[0, 1\right]$, denoting the gradient penalty~\cite{DBLP:conf/nips/GulrajaniAADC17}.

The loss function for training the code discriminator $C_{\eta}$ is:
\begin{equation}
    \begin{aligned}
        \mathcal{L}_{a3}&(\mathbf{x}, \hat{\mathbf{z}}_{rnd};\eta) = \\& C_{\eta}(\hat{\mathbf{z}}_{rec}) - C_{\eta}(\hat{\mathbf{z}}_{rnd}) + \lambda_4 \left(\|\nabla_{\tilde{\mathbf{z}}}C_{\eta}\left(\tilde{\mathbf{z}}\right) \|_{2} - 1\right)^{2},
    \end{aligned}
\end{equation}
where $\tilde{\mathbf{z}} = \epsilon''\hat{\mathbf{z}}_{rnd} + (1 - \epsilon'')\hat{\mathbf{z}}_{rec}, \epsilon''\in U\left[0, 1\right]$.
$\lambda_{1}, \lambda_{2}, \lambda_{3},$ and $\lambda_{4}$ are hyperparameters for balancing loss terms.

The generator and discriminator of $\alpha$-GAN share the same structures of the generator and discriminator of the GAN model, respectively.
The encoder is implemented based on the discriminator of the GAN model by changing the output dimension of the last FFN layer from 1 to 1024. 
The code discriminator is implemented as an FFN block with three linear layers.
We train the $\alpha$-GAN model with the Adam optimization as well.
After training, new synthetic segmentations can be generated by feeding the generator with noise vectors sampled from the prior distribution.

\subsection{The Fine-tuning Process}

The fine-tuning process aims to utilize the aforementioned dataset $\mathcal{X}_{fin} = \{(\mathbf{x}^{i}, \mathbf{v}^{i}, \mathbf{y}^{i})\}$ to train four conditional generative models, i.e., C-VAE, C-GAN, C-LDM, and CSegSynth.

\subsubsection{Conditional Variational Autoencoder (C-VAE)}

The general structure of the C-VAE model~\cite{DBLP:conf/nips/SohnLY15} is depicted in \cref{fig:structures}~(b), where the model consists of a conditional encoder, a conditional decoder, and a prior network.
Compared with the pre-trained VAE model, both the encoding of the segmentation $\mathbf{x}$ and the decoding of the sampled vector $\mathbf{z}^{(l)}$ are conditioned on the individual's feature vector $\mathbf{y}$, i.e., $q_{\phi}(\mathbf{z}|\mathbf{x}, \mathbf{y})$ and $p_{\theta}(\mathbf{x}|\mathbf{z}^{(l)}, \mathbf{y})$, respectively.
The conditional encoding is achieved by conditioning the input of the two FFN blocks in the original VAE encoder (cf. left bottom in \cref{fig:enc_dec}) on $\mathbf{y}$.
Specifically, denoting the original input of the two FFN blocks as $\mathbf{h}$, we add additional FFN layers, denoted as the function $\FFN_{e}(\cdot)$, to update $\mathbf{h}$ to $\mathbf{h} = \mathbf{h} + \FFN_{e}(\mathbf{y})$.
The conditional decoding is achieved by conditioning the sampled vector $\mathbf{z}^{(l)}$ on $\mathbf{y}$ in a similar way: $\mathbf{z}^{(l)} = \mathbf{z}^{(l)} + \FFN_{d}(\mathbf{y})$.
The prior network aims to estimate a prior distribution that is conditioned on $\mathbf{y}$, i.e., $p_{\theta}(\mathbf{z}|\mathbf{y})$.
We assume the prior distribution to be an isotropic multivariate Gaussian, i.e., $p_{\theta}(\mathbf{z}|\mathbf{y}) = \mathcal{N}(\mathbf{z}; \boldsymbol{\mu}, \boldsymbol{\sigma}^{2}\mathbf{I})$.
Accordingly, the prior network is implemented as two FFN blocks that compute $\boldsymbol{\mu}$ and $\boldsymbol{\sigma}^{2}$ based on $\mathbf{y}$: $\boldsymbol{\mu} = \FFN_{p1}(\mathbf{y})$ and $\boldsymbol{\sigma}^{2} = \FFN_{p2}(\mathbf{y})$.
The loss function regarding each real training segmentation $\mathbf{x}$ with its corresponding 
feature vector $\mathbf{y}$ and ground-truth volume $\mathbf{v}$ is defined as follows:
\begin{equation}
    \begin{aligned}
        \mathcal{L}_{cvae}&(\mathbf{x}, \mathbf{y}, \mathbf{v};\phi,\theta) =\\& \lambda_1 \KL\left(q_{\phi}(\mathbf{z}|\mathbf{x}, \mathbf{y})\|p_{\theta}(\mathbf{z}|\mathbf{y})\right) - \lambda_2 \frac{1}{L} \sum_{l=1}^{L} \log p_{\theta}(\mathbf{x}|\mathbf{z}^{(l)}, \mathbf{y})\\&+\lambda_3 \L1(\Vol(\hat{\mathbf{x}}), \mathbf{v}),
    \end{aligned}
\end{equation}
where $\phi$ and $\theta$ denote the parameters to learn, and $\KL\left(q_{\phi}(\mathbf{z}|\mathbf{x}, \mathbf{y})\|p_{\theta}(\mathbf{z}|\mathbf{y})\right)$ denotes the KL divergence between the encoded distribution $q_{\phi}(\mathbf{z}|\mathbf{x}, \mathbf{y})$ and the estimated prior $p_{\theta}(\mathbf{z}|\mathbf{y})$.
$\hat{\mathbf{x}}$ denotes a reconstruction of $\mathbf{x}$ generated by the decoder based on a random sample from $q_{\phi}(\mathbf{z}|\mathbf{x}, \mathbf{y})$.
$\Vol(\hat{\mathbf{x}})$ denotes the volume vector computed from $\hat{\mathbf{x}}$.
Furthermore, $\L1(\Vol(\hat{\mathbf{x}}), \mathbf{v})$ denotes the mean absolute error between $\Vol(\hat{\mathbf{x}})$ and $\mathbf{v}$.
$\lambda_{1}, \lambda_{2},$ and $\lambda_{3}$ are hyperparameters for balancing loss terms.
We used the Adam algorithm for training, after which, given a new individual with the feature vector $\hat{\mathbf{y}}$, we first use the prior network to estimate the conditional prior distribution $p_{\theta}(\mathbf{z}|\hat{\mathbf{y}})$.
Then, we sample a vector from $p_{\theta}(\mathbf{z}|\hat{\mathbf{y}})$ and use the decoder to generate $\hat{\mathbf{x}}$, i.e., $\hat{\mathbf{x}} = \argmax_{\mathbf{x}} p_{\theta}(\mathbf{x}|\hat{\mathbf{z}}, \hat{\mathbf{y}}), \hat{\mathbf{z}}\sim p_{\theta}(\mathbf{z}|\hat{\mathbf{y}})$, where $\hat{\mathbf{x}}$ is conditioned on $\hat{\mathbf{y}}$ and, therefore, specific to this individual.

\subsubsection{Conditional Generative Adversarial Network (C-GAN)}
\label{subsubsec:cgan}

We present an overview of the C-GAN structure in \cref{fig:structures}~(d), which is designed for the segmentation generation task of this paper based on the generic conditional generative adversarial nets~\cite{DBLP:journals/corr/MirzaO14}.
Compared with the pre-trained GAN model, cf. \cref{fig:structures}~(c), there are mainly two changes:
1) The generator $G_{\theta}$ becomes conditioned on the input feature vector $\mathbf{y}$, i.e., $\hat{\mathbf{x}} = G_{\theta}(\hat{\mathbf{z}}| \mathbf{y}), \hat{\mathbf{z}} \sim \mathcal{N}(\mathbf{z}; \mathbf{0}, \mathbf{I})$.
In this way, we can condition the synthetic segmentations on the input features and generate segmentations for specific individuals.
2) We compute the WM, GM, CSF, and background volumes in generated segmentations and use them in the training loss.
Specifically, we add the prediction error $\L1(\Vol(\hat{\mathbf{x}}), \mathbf{v})$ as an additional term of the generator's training loss, where $\Vol(\hat{\mathbf{x}})$ denotes computed volumes in $\hat{\mathbf{x}}$, and $\L1(\Vol(\hat{\mathbf{x}}), \mathbf{v})$ computes the mean absolute error with respect to $\mathbf{v}$.
The loss functions for the generator and discriminator are defined as follows:
\begin{equation}
\label{equ:cgan_gen}
\begin{aligned}
    \mathcal{L}_{cgen}&(\hat{\mathbf{z}}, \mathbf{y}, \mathbf{v};\theta) = \\&- D_{\omega}(\hat{\mathbf{x}}) + \lambda_1 \L1\left(\Vol\left(\hat{\mathbf{x}}\right), \mathbf{v}\right) \\&+ \lambda_2 \TML\left(\mathbf{v}, \Vol\left(\hat{\mathbf{x}}\right), \Vol\left(\hat{\mathbf{x}}'\right) \right),
\end{aligned}
\end{equation}
\begin{equation}
\begin{aligned}
    \mathcal{L}_{cdis}(\mathbf{x}, \mathbf{y};\omega) = D_{\omega}(\hat{\mathbf{x}}) - D_{\omega}(\mathbf{x}) + \lambda_3 (\|\nabla_{\tilde{\mathbf{x}}}D_{\omega}(\tilde{\mathbf{x}}) \|_{2} - 1)^{2},
\end{aligned}
\end{equation}
where $\hat{\mathbf{z}} \sim \mathcal{N}(\mathbf{z}; \mathbf{0}, \mathbf{I})$, $\hat{\mathbf{x}} = G_{\theta}(\hat{\mathbf{z}}| \mathbf{y})$, $\tilde{\mathbf{x}} = \epsilon \mathbf{x} + (1 - \epsilon) \hat{\mathbf{x}}$, $\epsilon \sim U\left[0, 1\right]$.
Please note that $\TML\left(\mathbf{v}, \Vol\left(\hat{\mathbf{x}}\right), \Vol\left(\hat{\mathbf{x}}'\right) \right)$ denotes a triplet margin loss~\cite{schroff2015facenet}, which uses $\mathbf{v}$ as the ``anchor,'' $\Vol\left(\hat{\mathbf{x}}\right)$ as the ``positive example'' that should be close to $\mathbf{v}$, and $\Vol\left(\hat{\mathbf{x}}'\right)$ as the ``negative example'' that should deviate from $\mathbf{v}$ by a margin compared to $\Vol\left(\hat{\mathbf{x}}\right)$.
Here, $\hat{\mathbf{x}}'$ denotes a random synthetic segmentation generated by the generator based on $\hat{\mathbf{z}}$ and $\mathbf{y} + \mathbf{y}_n$, where $\mathbf{y}_n$ has the same size of $\mathbf{y}$ but is sampled from a standard Gaussian.
$\lambda_1$, $\lambda_2$ and $\lambda_3$ are hyperparameters for balancing loss terms.

The discriminator shares the structure of the discriminator of the pre-trained GAN model.
In the conditional generator, we add an additional FFN layer, denoted as $\FFN_{cgan}(\cdot)$, on top of the original generator of the pre-trained GAN model.
The FFN layer conditions the input vector $\hat{\mathbf{z}}$ on $\mathbf{y}$, i.e., $\hat{\mathbf{z}} = \hat{\mathbf{z}} + \FFN_{cgan}(\mathbf{y})$.
The generator and discriminator are updated in separate steps based on the Adam algorithm.
After training, given a new individual with the feature vector $\hat{\mathbf{y}}$, the individual-specific segmentations can be generated as $\hat{\mathbf{x}} = G_{\theta}(\hat{\mathbf{z}}|\hat{\mathbf{y}}), \hat{\mathbf{z}}\sim \mathcal{N}(\mathbf{z}; \mathbf{0}, \mathbf{I})$.

\subsubsection{Conditional Latent Diffusion Model (C-LDM)}

An overview of the C-LDM model~\cite{DBLP:conf/cvpr/RombachBLEO22} is given in \cref{fig:structures}~(f).
Compared to LDM, cf. \cref{fig:structures}~(e), the denoising process of C-LDM is conditioned on the feature vector $\mathbf{y}$.
Specifically, the UNet module in the denoising process of the original LDM is augmented with a cross-attention mechanism~\cite{vaswani2017attention} that uses a transformation of $\mathbf{y}$ as ``keys'' and  ``values'' in the cross attention, as elaborated in \cite{DBLP:conf/cvpr/RombachBLEO22}.
Correspondingly, the loss function is updated as:
\begin{equation}
    \begin{aligned}
        \mathcal{L}_{cldm}(\mathbf{z}, \epsilon, t, \mathbf{y};\theta) = \|\epsilon - \epsilon_{\theta}(\mathbf{z_{t}}, t, \mathbf{y})\|_{2},
    \end{aligned}
\end{equation}
where $\epsilon_{\theta}(\mathbf{z_{t}}, t, \mathbf{y})$ denotes the output of the conditional UNet module at step $t$ given the latent representation of the current step, i.e., $\mathbf{z_{t}}$, and the feature vector $\mathbf{y}$.
Other notations are defined the same as \cref{equ:ldm}.
We used the Adam algorithm for training, after which, C-LDM can conduct the denoising of a random sample $\hat{\mathbf{z}}_{T}$ from $\mathcal{N}(\mathbf{z}; \mathbf{0}, \mathbf{I})$ with the condition of a given individual's feature vector and generate a synthetic segmentation $\hat{\mathbf{x}}$ that is specific to the individual, as depicted in \cref{fig:structures}~(f).

\subsubsection{Conditional Segmentation Synthesis Model (CSegSynth)}
\label{subsubsec:csegsynth}

An overview of the proposed CSegSynth model is given in \cref{fig:csegsynth_structures} (b), which consists of an encoder $E_{\phi}$, a code discriminator $C_{\eta}$, a prior network $P_{\psi}$, a generator $G_{\theta}$, and a discriminator $D_{\omega}$.
During training, given a sampled real segmentation $\mathbf{x}$.
The encoder $E_{\phi}$ first encodes $\mathbf{x}$ into a vector $\hat{\mathbf{z}}_{rec}$, which, on the one hand, is fed to the code discriminator $C_{\eta}$ to obtain a code ``score'' $C_{\eta}(\hat{\mathbf{z}}_{rec})$, and on the other hand, is fed to the generator $G_{\theta}$ to obtain a reconstruction of $\mathbf{x}$, i.e., $\hat{\mathbf{x}}_{rec}$.
Then, we sample a random vector $\hat{\mathbf{z}}_{rnd}$ from a prior distribution $\mathcal{N}(\mathbf{z}; \mathbf{0}, \mathbf{I})$.
The prior network $P_{\psi}$ conditions $\hat{\mathbf{z}}_{rnd}$ on the feature vector $\mathbf{y}$ and computes $\hat{\mathbf{z}}_{cnd}$, which, on the one hand, is also fed to the code discriminator, i.e., $C_{\eta}(\hat{\mathbf{z}}_{cnd})$, and on the other hand, is fed to the generator, obtaining a new synthetic segmentation $\hat{\mathbf{x}}_{cnd}$ that is conditioned on $\mathbf{y}$.
Lastly, the discriminator $D_{\omega}$ computes ``scores'' for both the reconstruction and the synthetic segmentation: $D_{\omega}(\hat{\mathbf{x}}_{rec})$ and $D_{\omega}(\hat{\mathbf{x}}_{cnd})$.
Based on these notations, we propose three training functions that learn the parameters of involved modules in separate steps:
\begin{equation}
\label{equ:csegsynth_ca1}
    \begin{aligned}
        \mathcal{L}_{c1}&(\mathbf{x}, \mathbf{y}, \mathbf{v}, \hat{\mathbf{z}}_{rnd};\phi, \theta, \psi) =\\ &- D_{\omega}(\hat{\mathbf{x}}_{cnd}) - D_{\omega}(\hat{\mathbf{x}}_{rec}) - \lambda_1 \left( C_{\eta}(\hat{\mathbf{z}}_{cnd}) + C_{\eta}(\hat{\mathbf{z}}_{rec})\right)\\& + \lambda_2 \left(\L1\left(\Vol(\hat{\mathbf{x}}_{cnd}), \mathbf{v}\right) + \L1\left(\Vol(\hat{\mathbf{x}}_{rec}), \mathbf{v}\right)\right) \\&+ \lambda_3 L_{BCE}(\hat{\mathbf{x}}_{rec}, \mathbf{x}) + \lambda_4 \TML\left(\mathbf{v}, \Vol\left(\hat{\mathbf{x}}_{cnd}\right), \Vol\left(\hat{\mathbf{x}}'_{cnd}\right)\right),
    \end{aligned}
\end{equation}
where $\phi$, $\theta$, and $\psi$ denote that this function is for training the encoder, the generator, and the prior network.
Minimizing $- D_{\omega}(\hat{\mathbf{x}}_{cnd}) - D_{\omega}(\hat{\mathbf{x}}_{rec})$ drives the generator to generate segmentations with higher quality based on outputs of both the prior network and the encoder.
The term $- \lambda_1 \left( C_{\eta}(\hat{\mathbf{z}}_{cnd}) + C_{\eta}(\hat{\mathbf{z}}_{rec})\right)$ improves the prior network and the encoder based on the judgment of the code discriminator.
The term $\lambda_2 \left(\L1\left(\Vol(\hat{\mathbf{x}}_{cnd}), \mathbf{v}\right) + \L1\left(\Vol(\hat{\mathbf{x}}_{rec}), \mathbf{v}\right)\right)$ regularizes the generated segmentations regarding the ground-truth volume $\mathbf{v}$.
The term $\lambda_3 L_{BCE}(\hat{\mathbf{x}}_{rec}, \mathbf{x})$ improves the generator based on voxel-wise binary cross entropy loss between $\mathbf{x}$ and the reconstruction $\hat{\mathbf{x}}_{rec}$.
$\TML\left(\mathbf{v}, \Vol\left(\hat{\mathbf{x}}_{cnd}\right), \Vol\left(\hat{\mathbf{x}}'_{cnd}\right)\right)$ denotes a triplet margin loss~\cite{schroff2015facenet}, which uses $\mathbf{v}$ as the ``anchor,'' $\Vol\left(\hat{\mathbf{x}}_{cnd}\right)$ as the ``positive example'' that should be close to $\mathbf{v}$, and $\Vol\left(\hat{\mathbf{x}}'_{cnd}\right)$ as the ``negative example'' that should deviate from $\mathbf{v}$ by a margin compared to $\Vol\left(\hat{\mathbf{x}}_{cnd}\right)$.
Here, $\hat{\mathbf{x}}'_{cnd}$ denotes a random synthetic segmentation generated by the prior network and the generator based on $\hat{\mathbf{z}}_{rnd}$ and $\mathbf{y} + \mathbf{y}_n$, where $\mathbf{y}_n$ has the same size of $\mathbf{y}$ and is sampled from a standard Gaussian.

\begin{equation}
    \begin{aligned}
        \mathcal{L}_{c2}&(\mathbf{x}, \mathbf{y}, \hat{\mathbf{z}}_{rnd};\omega) =\\ &D_{\omega}(\hat{\mathbf{x}}_{cnd}) + D_{\omega}(\hat{\mathbf{x}}_{rec}) - 2 D_{\omega}(\mathbf{x}) 
        + \\& \lambda_5 \left((\|\nabla_{\tilde{\mathbf{x}}}D_{\omega}(\tilde{\mathbf{x}}) \|_{2} - 1)^{2} + (\|\nabla_{\tilde{\mathbf{x}}'}D_{\omega}(\tilde{\mathbf{x}}') \|_{2} - 1)^{2}\right),
    \end{aligned}
\end{equation}
where $\omega$ denotes that this function updates the discriminator.
Minimizing $D_{\omega}(\hat{\mathbf{x}}_{cnd}) + D_{\omega}(\hat{\mathbf{x}}_{rec}) - 2 D_{\omega}(\mathbf{x})$ drives the discriminator to give real segmentations higher ``scores'' and give the reconstructed and synthetic ones lower ``scores,'' i.e., having a stronger discriminative capability.
The term $\lambda_5 \left((\|\nabla_{\tilde{\mathbf{x}}}D_{\omega}(\tilde{\mathbf{x}}) \|_{2} - 1)^{2} + (\|\nabla_{\tilde{\mathbf{x}}'}D_{\omega}(\tilde{\mathbf{x}}') \|_{2} - 1)^{2}\right)$ computes the gradient penalty~\cite{DBLP:conf/nips/GulrajaniAADC17} that regularizes the training. where $\tilde{\mathbf{x}} = \epsilon \mathbf{x} + (1 - \epsilon) \hat{\mathbf{x}}_{cnd}, \tilde{\mathbf{x}}' = \epsilon' \mathbf{x} + (1 - \epsilon') \hat{\mathbf{x}}_{rec}, \epsilon, \epsilon' \sim U\left[0, 1\right]$.
\begin{equation}
    \begin{aligned}
        \mathcal{L}_{c3}&(\mathbf{x}, \mathbf{y}, \hat{\mathbf{z}}_{rnd};\eta) =\\& C_{\eta}(\hat{\mathbf{z}}_{cnd}) + C_{\eta}(\hat{\mathbf{z}}_{rec}) - 2 C_{\eta}(\hat{\mathbf{z}}_{rnd})
         + \\& \lambda_6 \left( \left(\|\nabla_{\tilde{\mathbf{z}}}C_{\eta}(\tilde{\mathbf{z}}) \|_{2} - 1\right)^{2} + \left(\|\nabla_{\tilde{\mathbf{z}}'}C_{\eta}(\tilde{\mathbf{z}}') \|_{2} - 1\right)^{2} \right),
    \end{aligned}
\end{equation}
where $\eta$ denotes that this function updates the code discriminator.
Minimizing $C_{\eta}(\hat{\mathbf{z}}_{cnd}) + C_{\eta}(\hat{\mathbf{z}}_{rec}) - 2 C_{\eta}(\hat{\mathbf{z}}_{rnd})$ drives the code discriminator to have a stronger discriminative capability against the outputs of the prior network and encoder.
The last term also denotes a gradient penalty, where
$\tilde{\mathbf{z}} = \hat{\epsilon} \hat{\mathbf{z}}_{rnd} + (1 - \hat{\epsilon}) \hat{\mathbf{z}}_{cnd}, \tilde{\mathbf{z}}' = \hat{\epsilon}' \hat{\mathbf{z}}_{rnd} + (1 - \hat{\epsilon}') \hat{\mathbf{z}}_{rec},$ and $\hat{\epsilon}, \hat{\epsilon}'\sim U\left[0, 1\right]$.
Furthermore, $\lambda_{1}, \lambda_{2}, \lambda_{3}, \lambda_{4}, \lambda_{5}$ and $\lambda_6$ are hyperparameters for balancing loss terms.

The prior network is implemented as an FFN block with three linear layers, i.e., $\hat{\mathbf{z}}_{cnd} = \hat{\mathbf{z}}_{rnd} + \FFN_{prior}(\mathbf{y})$.
The other modules employ corresponding structures in $\alpha$-GAN.
We adopted the Adam optimization algorithm as well.
After training, given an individual with the feature vector $\hat{\mathbf{y}}$, we can utilize the prior network and the generator to conditionally generate an individual-specific segmentation $\hat{\mathbf{x}} = G_{\theta}(P_{\psi}(\hat{\mathbf{z}}|\hat{\mathbf{y}})), \hat{\mathbf{z}}\sim \mathcal{N}(\mathbf{z}; \mathbf{0}, \mathbf{I})$.

\subsection{Statistics}
In \cref{tab:correlations}, the Pearson correlation coefficients are solely reported as descriptive metrics.
No statistical tests were performed as the CSF volumes and C-LDM outputs do not satisfy the normality assumption according to Shapiro–Wilk tests~\cite{shapiro1965analysis}.
The other two non-parametric correlations---Spearman's $\rho$ and Kendall's $\tau$---are two-tailed and reported with exact p-values in the table.
The $n$ values are 46, i.e., the size of the CamCAN test set.
In \cref{fig:test_reg_traj}, the regression lines are only to illustrate the trend of the predicted volumes.
Each line represents a simple linear regression based on $n = 33$, with a 95\% confidence interval shown in the figure.

\section{Data Availability}

The AOMIC data that support the findings of this study are available in OpenNeuro (\href{https://openneuro.org/}{https://openneuro.org/}) with the identifier ``\href{https://openneuro.org/datasets/ds003097/versions/1.2.1}{doi:10.18112/openneuro.ds003097.v1.2.1}''.
The CamCAN data are available upon application from the Cambridge Centre for Ageing and Neuroscience (\href{https://cam-can.mrc-cbu.cam.ac.uk/}{https://cam-can.mrc-cbu.cam.ac.uk/}) at \href{https://camcan-archive.mrc-cbu.cam.ac.uk/dataaccess/}{https://camcan-archive.mrc-cbu.cam.ac.uk/dataaccess/}.
The exact numbers of the predicted WM and GM volumes in \cref{fig:test_reg_traj} are reported in the supplementary document (Supplementary Table 1).

\section{Code Availability}

The source code is publicly available at \href{https://github.com/ruijie-wang-uzh/CSegSynth}{https://github.com/ruijie-wang-uzh/CSegSynth}.
We also provide detailed information on the recommended computational resources, training configurations, and the runtime of each training process in the supplementary document.

\bibliographystyle{splncs04}
\bibliography{bibfile.bib}

\begin{thebibliography}{10}
\providecommand{\url}[1]{\texttt{#1}}
\providecommand{\urlprefix}{URL }
\providecommand{\doi}[1]{https://doi.org/#1}

\bibitem{ba2016layer}
Ba, J.L., Kiros, J.R., Hinton, G.E.: Layer normalization. arXiv preprint
  arXiv:1607.06450  (2016)

\bibitem{DBLP:journals/pami/Bond-TaylorLLW22}
Bond{-}Taylor, S., Leach, A., Long, Y., Willcocks, C.G.: Deep generative
  modelling: {A} comparative review of vaes, gans, normalizing flows,
  energy-based and autoregressive models. {IEEE} Trans. Pattern Anal. Mach.
  Intell.  \textbf{44}(11),  7327--7347 (2022)

\bibitem{burgmans2010multiple}
Burgmans, S., van Boxtel, M.P., Gronenschild, E., Vuurman, E., Hofman, P.,
  Uylings, H.B., Jolles, J., Raz, N.: Multiple indicators of age-related
  differences in cerebral white matter and the modifying effects of
  hypertension. Neuroimage  \textbf{49}(3),  2083--2093 (2010)

\bibitem{DBLP:journals/corr/abs-2211-02701}
Cardoso, M.J., Li, W., Brown, R., Ma, N., Kerfoot, E., Wang, Y., Murray, B.,
  Myronenko, A., Zhao, C., Yang, D., Nath, V., He, Y., Xu, Z., Hatamizadeh, A.,
  Zhu, W., Liu, Y., Zheng, M., Tang, Y., Yang, I., Zephyr, M., Hashemian, B.,
  Alle, S., Darestani, M.Z., Budd, C., Modat, M., Vercauteren, T., Wang, G.,
  Li, Y., Hu, Y., Fu, Y., Gorman, B., Johnson, H.J., Genereaux, B.W., Erdal,
  B.S., Gupta, V., Diaz{-}Pinto, A., Dourson, A., Maier{-}Hein, L., Jaeger,
  P.F., Baumgartner, M., Kalpathy{-}Cramer, J., Flores, M., Kirby, J.S.,
  Cooper, L.A.D., Roth, H.R., Xu, D., Bericat, D., Floca, R., Zhou, S.K.,
  Shuaib, H., Farahani, K., Maier{-}Hein, K.H., Aylward, S.R., Dogra, P.,
  Ourselin, S., Feng, A.: {MONAI:} an open-source framework for deep learning
  in healthcare. CoRR  \textbf{abs/2211.02701} (2022)

\bibitem{cattell1971abilities}
Cattell, R.: Abilities: Their structure, growth, and action (1971)

\bibitem{chen2019med3d}
Chen, S., Ma, K., Zheng, Y.: Med3d: Transfer learning for 3d medical image
  analysis. arXiv preprint arXiv:1904.00625  (2019)

\bibitem{DBLP:journals/titb/DorjsembePOX24}
Dorjsembe, Z., Pao, H., Odonchimed, S., Xiao, F.: Conditional diffusion models
  for semantic 3d brain {MRI} synthesis. {IEEE} J. Biomed. Health Informatics
  \textbf{28}(7),  4084--4093 (2024)

\bibitem{esser2021taming}
Esser, P., Rombach, R., Ommer, B.: Taming transformers for high-resolution
  image synthesis. In: Proceedings of the IEEE/CVF conference on computer
  vision and pattern recognition. pp. 12873--12883 (2021)

\bibitem{DBLP:journals/mia/FernandezPBGTVC24}
Fernandez, V., Pinaya, W.H.L., Borges, P., Graham, M.S., Tudosiu, P.,
  Vercauteren, T., Cardoso, M.J.: Generating multi-pathological and multi-modal
  images and labels for brain {MRI}. Medical Image Anal.  \textbf{97},  103278
  (2024)

\bibitem{fletcher2013loss}
Fletcher, E., Raman, M., Huebner, P., Liu, A., Mungas, D., Carmichael, O.,
  DeCarli, C.: Loss of fornix white matter volume as a predictor of cognitive
  impairment in cognitively normal elderly individuals. JAMA neurology
  \textbf{70}(11),  1389--1395 (2013)

\bibitem{gencc2018diffusion}
Gen{\c{c}}, E., Fraenz, C., Schl{\"u}ter, C., Friedrich, P., Hossiep, R.,
  Voelkle, M.C., Ling, J.M., G{\"u}nt{\"u}rk{\"u}n, O., Jung, R.E.: Diffusion
  markers of dendritic density and arborization in gray matter predict
  differences in intelligence. Nature communications  \textbf{9}(1), ~1905
  (2018)

\bibitem{gretton2012kernel}
Gretton, A., Borgwardt, K.M., Rasch, M.J., Sch{\"o}lkopf, B., Smola, A.: A
  kernel two-sample test. The Journal of Machine Learning Research
  \textbf{13}(1),  723--773 (2012)

\bibitem{DBLP:conf/nips/GulrajaniAADC17}
Gulrajani, I., Ahmed, F., Arjovsky, M., Dumoulin, V., Courville, A.C.: Improved
  training of wasserstein gans. In: Guyon, I., von Luxburg, U., Bengio, S.,
  Wallach, H.M., Fergus, R., Vishwanathan, S.V.N., Garnett, R. (eds.) Advances
  in Neural Information Processing Systems 30: Annual Conference on Neural
  Information Processing Systems 2017, December 4-9, 2017, Long Beach, CA,
  {USA}. pp. 5767--5777 (2017)

\bibitem{heusel2017gans}
Heusel, M., Ramsauer, H., Unterthiner, T., Nessler, B., Hochreiter, S.: Gans
  trained by a two time-scale update rule converge to a local nash equilibrium.
  Advances in neural information processing systems  \textbf{30} (2017)

\bibitem{heyn2025differential}
Heyn, S.A., Keding, T.J., Cisler, J., McLaughlin, K., Herringa, R.J.:
  Differential gray matter correlates and machine learning prediction of abuse
  and internalizing psychopathology in adolescent females. Scientific Reports
  \textbf{15}(1), ~651 (2025)

\bibitem{ho2020denoising}
Ho, J., Jain, A., Abbeel, P.: Denoising diffusion probabilistic models.
  Advances in neural information processing systems  \textbf{33},  6840--6851
  (2020)

\bibitem{hodges1994neurological}
Hodges, J.R.: Neurological aspects of dementia and normal aging. Dementia and
  Normal Aging pp. 118--129 (1994)

\bibitem{ibrahim2012cost}
Ibrahim, R., Samian, S., Mazli, M., Amrizal, M., Aljunid, S.M.: Cost of
  magnetic resonance imaging (mri) and computed tomography (ct) scan in ukmmc.
  BMC Health Services Research  \textbf{12}(Suppl 1), ~P11 (2012)

\bibitem{DBLP:journals/csur/JabbarLO22}
Jabbar, A., Li, X., Omar, B.: A survey on generative adversarial networks:
  Variants, applications, and training. {ACM} Comput. Surv.  \textbf{54}(8),
  157:1--157:49 (2022)

\bibitem{jiang2023white}
Jiang, J., Bruss, J., Lee, W.T., Tranel, D., Boes, A.D.: White matter
  disconnection of left multiple demand network is associated with post-lesion
  deficits in cognitive control. Nature communications  \textbf{14}(1), ~1740
  (2023)

\bibitem{DBLP:journals/corr/KingmaB14}
Kingma, D.P., Ba, J.: Adam: {A} method for stochastic optimization. In: Bengio,
  Y., LeCun, Y. (eds.) 3rd International Conference on Learning
  Representations, {ICLR} 2015, San Diego, CA, USA, May 7-9, 2015, Conference
  Track Proceedings (2015)

\bibitem{DBLP:journals/corr/KingmaW13}
Kingma, D.P., Welling, M.: Auto-encoding variational bayes. In: Bengio, Y.,
  LeCun, Y. (eds.) 2nd International Conference on Learning Representations,
  {ICLR} 2014, Banff, AB, Canada, April 14-16, 2014, Conference Track
  Proceedings (2014)

\bibitem{DBLP:conf/miccai/KwonHK19}
Kwon, G., Han, C., Kim, D.: Generation of 3d brain {MRI} using auto-encoding
  generative adversarial networks. In: Shen, D., Liu, T., Peters, T.M., Staib,
  L.H., Essert, C., Zhou, S., Yap, P., Khan, A.R. (eds.) Medical Image
  Computing and Computer Assisted Intervention - {MICCAI} 2019 - 22nd
  International Conference, Shenzhen, China, October 13-17, 2019, Proceedings,
  Part {III}. Lecture Notes in Computer Science, vol. 11766, pp. 118--126.
  Springer (2019)

\bibitem{lavanga2023virtual}
Lavanga, M., Stumme, J., Yalcinkaya, B.H., Fousek, J., Jockwitz, C., Sheheitli,
  H., Bittner, N., Hashemi, M., Petkoski, S., Caspers, S., et~al.: The virtual
  aging brain: Causal inference supports interhemispheric dedifferentiation in
  healthy aging. NeuroImage  \textbf{283},  120403 (2023)

\bibitem{DBLP:journals/cmpb/LiuDBSYB23}
Liu, Y., Dwivedi, G., Boussa{\"{\i}}d, F., Sanfilippo, F.M., Yamada, M.,
  Bennamoun, M.: Inflating 2d convolution weights for efficient generation of
  3d medical images. Comput. Methods Programs Biomed.  \textbf{240},  107685
  (2023)

\bibitem{DBLP:conf/cvpr/0003MWFDX22}
Liu, Z., Mao, H., Wu, C., Feichtenhofer, C., Darrell, T., Xie, S.: A convnet
  for the 2020s. In: {IEEE/CVF} Conference on Computer Vision and Pattern
  Recognition, {CVPR} 2022, New Orleans, LA, USA, June 18-24, 2022. pp.
  11966--11976. {IEEE} (2022). \doi{10.1109/CVPR52688.2022.01167}

\bibitem{DBLP:conf/iclr/LoshchilovH19}
Loshchilov, I., Hutter, F.: Decoupled weight decay regularization. In: 7th
  International Conference on Learning Representations, {ICLR} 2019, New
  Orleans, LA, USA, May 6-9, 2019. OpenReview.net (2019)

\bibitem{mei2022radimagenet}
Mei, X., Liu, Z., Robson, P.M., Marinelli, B., Huang, M., Doshi, A., Jacobi,
  A., Cao, C., Link, K.E., Yang, T., et~al.: Radimagenet: an open radiologic
  deep learning research dataset for effective transfer learning. Radiology:
  Artificial Intelligence  \textbf{4}(5),  e210315 (2022)

\bibitem{DBLP:journals/corr/MirzaO14}
Mirza, M., Osindero, S.: Conditional generative adversarial nets. CoRR
  \textbf{abs/1411.1784} (2014)

\bibitem{DBLP:conf/miccai/PinayaTDCFNOC22}
Pinaya, W.H.L., Tudosiu, P., Dafflon, J., Costa, P.F.D., Fernandez, V., Nachev,
  P., Ourselin, S., Cardoso, M.J.: Brain imaging generation with latent
  diffusion models. In: Mukhopadhyay, A., {\"{O}}ks{\"{u}}z, I., Engelhardt,
  S., Zhu, D., Yuan, Y. (eds.) Deep Generative Models - Second {MICCAI}
  Workshop, {DGM4MICCAI} 2022, Held in Conjunction with {MICCAI} 2022,
  Singapore, September 22, 2022, Proceedings. Lecture Notes in Computer
  Science, vol. 13609, pp. 117--126. Springer (2022)

\bibitem{rodd2010functional}
Rodd, J.M., Longe, O.A., Randall, B., Tyler, L.K.: The functional organisation
  of the fronto-temporal language system: evidence from syntactic and semantic
  ambiguity. Neuropsychologia  \textbf{48}(5),  1324--1335 (2010)

\bibitem{DBLP:conf/cvpr/RombachBLEO22}
Rombach, R., Blattmann, A., Lorenz, D., Esser, P., Ommer, B.: High-resolution
  image synthesis with latent diffusion models. In: {IEEE/CVF} Conference on
  Computer Vision and Pattern Recognition, {CVPR} 2022, New Orleans, LA, USA,
  June 18-24, 2022. pp. 10674--10685. {IEEE} (2022).
  \doi{10.1109/CVPR52688.2022.01042}

\bibitem{ronneberger2015u}
Ronneberger, O., Fischer, P., Brox, T.: U-net: Convolutional networks for
  biomedical image segmentation. In: Medical image computing and
  computer-assisted intervention--MICCAI 2015: 18th international conference,
  Munich, Germany, October 5-9, 2015, proceedings, part III 18. pp. 234--241.
  Springer (2015)

\bibitem{DBLP:journals/corr/RoscaLWM17}
Rosca, M., Lakshminarayanan, B., Warde{-}Farley, D., Mohamed, S.: Variational
  approaches for auto-encoding generative adversarial networks. CoRR
  \textbf{abs/1706.04987} (2017)

\bibitem{rumelhart1986learning}
Rumelhart, D.E., Hinton, G.E., Williams, R.J.: Learning internal
  representations by error propagation, parallel distributed processing,
  explorations in the microstructure of cognition, ed. de rumelhart and j.
  mcclelland. vol. 1. 1986. Biometrika  \textbf{71}(599-607), ~6 (1986)

\bibitem{schroff2015facenet}
Schroff, F., Kalenichenko, D., Philbin, J.: Facenet: A unified embedding for
  face recognition and clustering. In: Proceedings of the IEEE conference on
  computer vision and pattern recognition. pp. 815--823 (2015)

\bibitem{shafto2014cambridge}
Shafto, M.A., Tyler, L.K., Dixon, M., Taylor, J.R., Rowe, J.B., Cusack, R.,
  Calder, A.J., Marslen-Wilson, W.D., Duncan, J., Dalgleish, T., et~al.: The
  cambridge centre for ageing and neuroscience (cam-can) study protocol: a
  cross-sectional, lifespan, multidisciplinary examination of healthy cognitive
  ageing. BMC neurology  \textbf{14},  1--25 (2014)

\bibitem{shapiro1965analysis}
Shapiro, S.S., Wilk, M.B.: An analysis of variance test for normality (complete
  samples). Biometrika  \textbf{52}(3-4),  591--611 (1965)

\bibitem{snoek2021amsterdam}
Snoek, L., van~der Miesen, M.M., Beemsterboer, T., Van Der~Leij, A., Eigenhuis,
  A., Steven~Scholte, H.: The amsterdam open mri collection, a set of
  multimodal mri datasets for individual difference analyses. Scientific data
  \textbf{8}(1), ~85 (2021)

\bibitem{DBLP:conf/nips/SohnLY15}
Sohn, K., Lee, H., Yan, X.: Learning structured output representation using
  deep conditional generative models. In: Cortes, C., Lawrence, N.D., Lee,
  D.D., Sugiyama, M., Garnett, R. (eds.) Advances in Neural Information
  Processing Systems 28: Annual Conference on Neural Information Processing
  Systems 2015, December 7-12, 2015, Montreal, Quebec, Canada. pp. 3483--3491
  (2015)

\bibitem{DBLP:journals/titb/SunCXGYB22}
Sun, L., Chen, J., Xu, Y., Gong, M., Yu, K., Batmanghelich, K.: Hierarchical
  amortized {GAN} for 3d high resolution medical image synthesis. {IEEE} J.
  Biomed. Health Informatics  \textbf{26}(8),  3966--3975 (2022)

\bibitem{taylor2017cambridge}
Taylor, J.R., Williams, N., Cusack, R., Auer, T., Shafto, M.A., Dixon, M.,
  Tyler, L.K., Henson, R.N., et~al.: The cambridge centre for ageing and
  neuroscience (cam-can) data repository: Structural and functional mri, meg,
  and cognitive data from a cross-sectional adult lifespan sample. neuroimage
  \textbf{144},  262--269 (2017)

\bibitem{thompson2020tracking}
Thompson, D.K., Matthews, L.G., Alexander, B., Lee, K.J., Kelly, C.E., Adamson,
  C.L., Hunt, R.W., Cheong, J.L., Spencer-Smith, M., Neil, J.J., et~al.:
  Tracking regional brain growth up to age 13 in children born term and very
  preterm. Nature communications  \textbf{11}(1), ~696 (2020)

\bibitem{vaswani2017attention}
Vaswani, A., Shazeer, N., Parmar, N., Uszkoreit, J., Jones, L., Gomez, A.N.,
  Kaiser, L., Polosukhin, I.: Attention is all you need. Advances in neural
  information processing systems  \textbf{30} (2017)

\bibitem{wang2024virtual}
Wang, H.E., Triebkorn, P., Breyton, M., Dollomaja, B., Lemarechal, J.D.,
  Petkoski, S., Sorrentino, P., Depannemaecker, D., Hashemi, M., Jirsa, V.K.:
  Virtual brain twins: from basic neuroscience to clinical use. National
  Science Review  \textbf{11}(5),  nwae079 (2024)

\bibitem{wang2004image}
Wang, Z., Bovik, A.C., Sheikh, H.R., Simoncelli, E.P.: Image quality
  assessment: from error visibility to structural similarity. IEEE transactions
  on image processing  \textbf{13}(4),  600--612 (2004)

\bibitem{white2022data}
White, T., Blok, E., Calhoun, V.D.: Data sharing and privacy issues in
  neuroimaging research: Opportunities, obstacles, challenges, and monsters
  under the bed. Human Brain Mapping  \textbf{43}(1),  278--291 (2022)

\bibitem{DBLP:journals/tmi/YuZWSFB19}
Yu, B., Zhou, L., Wang, L., Shi, Y., Fripp, J., Bourgeat, P.: Ea-gans:
  Edge-aware generative adversarial networks for cross-modality {MR} image
  synthesis. {IEEE} Trans. Medical Imaging  \textbf{38}(7),  1750--1762 (2019)

\bibitem{zhang2018unreasonable}
Zhang, R., Isola, P., Efros, A.A., Shechtman, E., Wang, O.: The unreasonable
  effectiveness of deep features as a perceptual metric. In: Proceedings of the
  IEEE conference on computer vision and pattern recognition. pp. 586--595
  (2018)

\bibitem{zhang2008discrete}
Zhang, W., Luck, S.J.: Discrete fixed-resolution representations in visual
  working memory. Nature  \textbf{453}(7192),  233--235 (2008)

\bibitem{zhang2001segmentation}
Zhang, Y., Brady, M., Smith, S.: Segmentation of brain mr images through a
  hidden markov random field model and the expectation-maximization algorithm.
  IEEE transactions on medical imaging  \textbf{20}(1),  45--57 (2001)

\end{thebibliography}

\section*{Acknowledgments}

This work was partially funded by the University Research Priority Program ``Dynamics of Healthy Aging'' at the University of Zurich and the Swiss National Science Foundation through project MediaGraph (contract no.\ 202125).
Data collection and sharing for this project was provided by the Cambridge Centre for Ageing and Neuroscience (CamCAN). CamCAN funding was provided by the UK Biotechnology and Biological Sciences Research Council (grant number BB/H008217/1), together with support from the UK Medical Research Council and University of Cambridge, UK.

\section*{Author Contributions}
R.W., L.R., S.M., C.R., M.M., and A.B. contributed to the proposal and conceptualization of the research question and general idea of this work.
R.W. and L.R. contributed to the design, implementation, training, and evaluation of deep generative models.
S.M., C.R., M.M., and A.B. contributed to the design of experiments and analysis of experimental results.
R.W. contributed to the writing of the whole paper.
S.M. contributed to the writing of the discussion section.
L.R., S.M., C.R., M.M., and A.B. contributed to the proofreading and revision of the whole paper.
Additionally, M.M. and A.B. contributed to the data and resource acquisition as well as the planning and coordination of the whole work.

\section*{Competing interests}
We declare no competing interests.
\\
\textbf{Correspondence} and material requests should be addressed to Ruijie Wang.

\end{document}